\documentclass[a4paper,11pt]{article}

\usepackage[dvips]{graphicx}
\usepackage[english]{babel}
\usepackage[latin1]{inputenc}
\usepackage{amssymb}
\usepackage{amsthm}
\usepackage[all]{xy}
\usepackage{url}

\usepackage{cite}
\usepackage{stfloats}

\def\rien{\rule{0pt}{0pt}}

\newcommand*{\bel}{\mathrm{Bel}}

\makeatletter
\def\overbracket#1{\mathop{\vbox{\ialign{##\crcr\noalign{\kern3\p@}
\downbracketfill\crcr\noalign{\kern3\p@\nointerlineskip}
$\hfil\displaystyle{#1}\hfil$\crcr}}}\limits}
\def\underbracket#1{\mathop{\vtop{\ialign{##\crcr
$\hfil\displaystyle{#1}\hfil$\crcr\noalign{\kern3\p@\nointerlineskip}
\upbracketfill\crcr\noalign{\kern3\p@}}}}\limits}
\def\overparenthesis#1{\mathop{\vbox{\ialign{##\crcr\noalign{\kern3\p@}
\downparenthfill\crcr\noalign{\kern3\p@\nointerlineskip}
$\hfil\displaystyle{#1}\hfil$\crcr}}}\limits}
\def\underparenthesis#1{\mathop{\vtop{\ialign{##\crcr
$\hfil\displaystyle{#1}\hfil$\crcr\noalign{\kern3\p@\nointerlineskip}
\upparenthfill\crcr\noalign{\kern3\p@}}}}\limits}
\def\downparenthfill{$\m@th\braceld\leaders\vrule\hfill\bracerd$}
\def\upparenthfill{$\m@th\bracelu\leaders\vrule\hfill\braceru$}
\def\upbracketfill{$\m@th\makesm@sh{\llap{\vrule\@height3\p@\@width.7\p@}}%
\leaders\vrule\@height.7\p@\hfill
\makesm@sh{\rlap{\vrule\@height3\p@\@width.7\p@}}$}
\def\downbracketfill{$\m@th
\makesm@sh{\llap{\vrule\@height.7\p@\@depth2.3\p@\@width.7\p@}}%
\leaders\vrule\@height.7\p@\hfill
\makesm@sh{\rlap{\vrule\@height.7\p@\@depth2.3\p@\@width.7\p@}}$}
\makeatother

\def\rien{\rule{0pt}{0pt}}
\def\Rset{\mathrm{I\!R}}

\def\everTrue{\Omega}
\def\everFalse{\emptyset}
\def\notsharp{\not{\!\lhd\,}}

\def\oplusconj{\oplus_{\cap}} 
\def\oplusDS{\oplus_{\mathrm{DS}}} 
\def\oplusPCRv{\oplus_{\mathrm{PCR5}}}
\def\oplusEMR{\oplus_{\mathrm{H}}}

\def\equivclass#1{{\left\lfloor#1\right\rfloor}}
\def\setvee{\mathbf{v}}
\def\modelof#1{\overline{#1}}
\def\modelalt#1{\underline{#1}}
\def\xModel{p}
\def\YModel{E}
\def\simplemodal{{{\mathcal{L}}_{C[C]}}}

\newtheorem{property}{Property}
\newtheorem{lemma}[property]{Lemma}
\newtheorem{corollary}[property]{Corollary}
\newtheorem{definition}[property]{Definition}
\newtheorem{coroldef}[property]{Corollary of the definition}
\newtheorem{example}[property]{Example}

\begin{document}

 \setlength{\baselineskip}{1.5\baselineskip}

\title{An Interpretation of Belief Functions by means of a Probabilistic Multi-modal Logic}
\author{%
Fr\'ed\'eric Dambreville\\
ENSTA Bretagne, team STIC/REMS\\
2 rue François Verny\\
29806 Brest Cedex 9\\
{\tt Email: submit@fredericdambreville.com}}

\maketitle

\selectlanguage{english}

\begin{abstract}
While belief functions may be seen formally as a generalization of probabilistic distributions, the question of the interactions between belief functions and probability is still an issue in practice.
This question is difficult, since the contexts of use of these theory are notably different and the semantics behind these theories are not exactly the same.
A prominent issue is increasingly regarded by the community, that is the management of the conflicting information.  
Recent works have introduced new rules for handling the conflict redistribution while combining belief functions.
The notion of conflict, or its cancellation by an hypothesis of open world, seems by itself to prevent a direct interpretation of belief function in a probabilistic framework.
This paper addresses the question of a probabilistic interpretation of belief functions.
It first introduces and implements a theoretically grounded rule, which is in essence an adaptive conjunctive rule.
It is shown, how this rule is derived from a logical interpretation of the belief functions by means of a probabilistic multimodal logic;
in addition, a concept of source independence is introduced, based on a principle of entropy maximization.
\\\rien
\end{abstract}

\noindent
{\bf Keywords:} Belief Functions, Modal Logic, Probability, Entropy maximization

%

%
%
\section{Introduction}
\label{DSmTb2:Dmb:Sec:1}
In \cite{Dempster1967,dempster2}, A. Dempster introduced an elegant modelling of imprecise probabilistic beliefs, which formally implies a generalisation of probability, and a process for infering combinations of these beliefs.
This seminal work completed by G. Shafer\cite{Shafer1976} initiated the theory of belief functions for manipulating uncertain and imprecise information.
However, while belief functions may be seen formally as a generalization of probabilistic distributions, the question of the interactions between belief functions and probability is still an issue in practice.
Indeed, the contexts of use of these theory are notably different implying slightly different semantics in background.
On the one hand, belief functions may deal with non technical sources of information, and as a consequence are also confronted inherently to conflictual information.
On the other hand, the contexts of use of probability and bayesian reasonning are more technical and controlled, and prevent, for the most part, the issues of conflictual information.
It is often necessary to map the belief information to the probabilistic information for actual applications.
As will be shown subsequently, some issue for a probabilistic interpretation of belief function is implied by the notion of conflict between the sources, which is a product of the conjunctive combination rule for fusing belief function.
The conflict and the management of the conflict in the combination rules are topics of increasing interest for the community
Recent works have proposed evolutions of the historical rule of Dempster and Shafer (DS) for managing the conflict; Lefevre \emph{and al} explicitely introduced this issue \cite{Lefevre2002}, and many rule constructions followed \cite{Smarandache2005, Florea2006, Martin2006, Smarandache2006, Denoeux2006, martin2007}.
Especially, Florea, Jousselme and al\cite{Florea2006} proposed a family of rules which is adaptive with the conflict level.
In this case, there is an important idea: the redistribution policy is now changing automatically as a function of the conflict.
But this concern for the conflict is not new, and are for example ingredients of the famous rules of Dubois\&Prade \cite{Dubois86,Dubois88} or Yager \cite{Yager1987}.
The notion of conflict may be cancelled by an hypothesis of open world \cite{smets}, but this open world hypothesis does not map to the common formalisms of probability, and have to be interpreted.
By the way, the open world hypothesis does not prevent any practical consideration of the conflict \cite{Smets07} implied by the interpretation of the belief put on the empty set. 
\\[5pt]
This paper addresses the question of a probabilistic interpretation of belief functions.
It is the continuation of reflexions done in \cite{DSmTb2:dmb,F2K7:dmb:emr}, which introduced and implemented a theoretically grounded rule, obtained by applying a principle of sources independence based entropy maximization.
This rule is in essence an adaptive conjunctive rule, where the conflict redistribution is automatically derived from the considered independence principle. 
It is shown that this rule identifies to the conjunctive rule in cases without conflict.
This rule is an introduction to the definition and study of a multimodal logic for modelling beliefs, since it is indeed derived from this logic by applying the independence principle.
We point out by means of the axiomatizations how the conflict may be interpreted from the logical level.
\\[5pt]
This paper is divided in two parts.
The first part introduces a methodology for computing combination rules, on the basis of an entropic principle for defining the independence between the sources.
It is started by a general discussion about belief functions in section~\ref{DSmTb2:Dmb:Sec:2}.
Section~\ref{DSmTb2:Dmb:Sec:3} settles a new methodology for deriving the combination rule, based on an entropic notion of sensor independence.
Then, section~\ref{DSmTb2:Dmb:Sec:4} discusses about the properties of the new rule and the implementations.
A small example is considered as an illustration, but since the purpose of this paper is mainly theoretical, these experimental aspects are not developped.
The second part exposes the logical fundaments of our methodology.
Section~\ref{logpart:sect:1} provides some logical backgrounds and introduces the definition of the multimodal logic.
In section~\ref{logpart:sect:3}, it is shown how a logical combination rule is derivable from the logic.
In section~\ref{logpart:sect:2}, a non trivial model is defined, and properties concerning the probabilistic extension of the logic are deduced.
Belief combination rules are then derived theoretically from the logical rule, by applying a maximization of the entropic information.
Section~\ref{DSmTb2:Dmb:Sec:6} concludes.
\part{A combination rule based on the entropy maximization principle}
\label{beliefpart}
\section{Overview}
\label{DSmTb2:Dmb:Sec:2}
%
\subsection{Frame of discernment}
%
In order to manipulate the information, it is necessary to have an algebraic representation of the events of the system.
Various structures~\cite{DSmTBook1,besnard1996} have been introduced recently,
and in the most general case, distributive lattices may be used for this algebraic representation~\cite{besnard1996}.
In this paper, however, we will restrict our reflexion to Boolean algebras, and especially to power sets, which are the historical and prominent structures used by the belief function community.
%
\begin{definition}[frame of discernment]\label{definition:1}
It is given a finite Boolean algabra $\mathbf{B}=(G,\cap,\cup,\sim,\everFalse, \everTrue)$ and a subset $\Theta\subset G$ of propositions, called \emph{frame of discernment}, which generates $G$.\footnote{Which means that $G$ is the set of all propositions derived from $\emptyset, \Omega,$ the elements of $\Theta$ and the operators $\cap,\cup,\sim$\,.}
\end{definition}
The frame of discernment $\Theta$ is the set of basical known events of our model, and the Boolean algebra $\mathbf{B}$ describes the logical interaction between the basical event of $\Theta$.
Subsequently, the notations $\mathbf{B}^{\Theta}\stackrel{\Delta}{=}\mathbf{B}$ and $G^{\Theta}\stackrel{\Delta}{=}G$ are used,
 in order to mark the structures with their frames of discernment.
%
\\[5pt]
While this paper addresses the general case of Boolean structures, it is known that power sets are sufficiently versatile structures, which may be equivalent representations of any finite Boolean algebra, and may as well include as substructures some lattice structures, like hyperpower sets~\cite{DSmTBook1},.
Thus, a fundamental example of Boolean structure $\mathbf{B}^{\Theta}$ is the example of power sets.
\begin{example}[power set]\label{example:0:1} 
\begin{itemize}
\item The frame of discernment identifies to singletons of a set $\Omega$, so that $\Theta =\bigl\{\{\omega\}\big/\omega\in\Omega\bigr\}$,
\item The Boolean algebra $\mathbf{B}^\Theta$ identifies to the algebra of the subsets of $\Omega$, so that $G^\Theta=2^\Omega$, and $\cap,\cup,\sim, \everFalse, \everTrue$ are respectiveley the set intersection, set union, set complement, the empty set and the entire set $\Omega$\,.
\end{itemize}
\end{example}
\subsection{Belief}
\subsubsection{Basic belief asignment}
The information from different sources or expert is not reducible to the only logical representation.
Information of real life are typically uncertain and imprecise and come with an uncertainty degree, or belief. 
In belief function theory, the basic belief assignment (bba) is a belief density, describing the information intrinsic to each proposition.
This intrinsic information is the fraction of information of the proposition which is not inherited from connex propositions. 
There are two definitions of bba related respectively to an \emph{open world} or a \emph{closed world} hypothesis.
\begin{definition}[basic belief assignment]\label{definition:2}
Being given a finite Boolean algebra $\mathbf{B}^\Theta=(G^\Theta,\cap,\cup,\sim,\everFalse, \everTrue)$,
a basic belief assignment $m$ to $G^\Theta$ is a real valued function defined over $G^\Theta$ such that:
$$
\sum_{X\in G^\Theta}m(X)=1
\quad\mbox{and}\quad m\ge0
\;,
$$
and:
\begin{description}
\item[\rien\quad Open world hypothesis:] No other constraint,
\item[\rien\quad Closed world hypothesis:] $m(\everFalse)=0$\,.
\end{description}
\end{definition}
The theory of Transferable Belief Model, as defined by Smets~\cite{smets}, implements typically the open world hypothesis.
%
While the bba describes the information intrinsic to each proposition, the \emph{full} belief of a proposition is potentially the compilation of the basic beliefs connex to this proposition.
There is a pessimistic bound and an optimistic bound for the belief of a propostion, which are the credibility and the plausibility.
%
\begin{definition}[credibilty function]\label{definition:3}
The cedibility $\bel$ related to a bba $m$ is defined for any $X\in G^\Theta$ by:
\begin{equation}
\label{f2k5:DSmTcont:Eq:1}
\bel(X)=\sum_{Y\in G^\Theta:\everFalse\subsetneq Y\subset X}m(Y)
=\sum_{Y\in G^\Theta:Y\subset X}m(Y)-m(\everFalse)\;.
\end{equation}
\end{definition}
The credibility is a pessimistic bound,
only talking into account the belief of the propositions known with certainty.
%
%
%
\begin{definition}[plausibility function]\label{definition:4}
The plausibility is defined by $\mathrm{Pl}(X)=1-\bel(\sim X)$
for any $X\in G^\Theta$\,.
\end{definition}
%
\subsection{Combination rules}
\label{DSmTb2:Dmb:CH2:section:FusionRule}
A main contribution of the theory of belief functions consists in the combination rules.
It is assumed that two or more sources of information are providing a
viewpoint about the universe.
These viewpoints are described by specific bba for each source.
The question then is to make a unique representation of the information,
\emph{i.e.} a unique bba, from these several bbas.

\subsubsection{Conjunctive rule}
The conjunctive rule belief functions is a rule of reference for combining belief functions with the \emph{open world hypothesis}.
It is applied for fusing independent sources of information.
\begin{definition}[conjunctive combination]\label{definition:5}
Let $m_1$ and $m_2$ be $2$ bbas over $G^\Theta$\,.
The conjunctive combination $m_1\oplusconj m_2$ of $m_1$ and $m_2$ is defined by: 
\begin{equation}\label{dmb:dsmtb2:ch2:rule:eq:2}
m_1\oplusconj m_2(X)=\sum_{Y_1,Y_2\in G^\Theta : Y1\cap Y_2=X}m_1(Y_1) m_2(Y_2)\;.
\end{equation}
\end{definition}
It is proved that $m_1\oplusconj m_2$ is a bba when $m_1$ and $m_2$ are bbas.
However, the quantity $m_1\oplusconj m_2(\everFalse)$\,, called \emph{degree of conflict}, is not necessary zero, even if both $m_1$ and $m_2$ meet the constraint $m_1(\everFalse)=m_2(\everFalse)=0$.
Thus, $\oplusconj$ is an acceptable combination rule when the open world hypothesis is made, like in the TBM, but it is not acceptable in general when the closed world hypothesis is done.
However, there are classes of bbas for which the conjunctive rule is admissible for the closed world hypothesis, that is bbas restricted to Boolean substructures for which contradiction $\everFalse$ cannot be derived from the conjunction $\cap$.
In~\cite{dezert,DSmTBook1}, Dezert and Smarandache proposed such a structure, the hyperpower set.
On such structure, the conjunctive rule does not generate conflict and is admissible even with the closed world hypothesis. 
%
%
\subsubsection{Conflict management with the closed world hypothesis}
In the general case, the conjunctive rule generates a degree of conflict, which is not complient with the closed world hypothesis.
While many rules are based on the conjunctive rule, a management of the conflict is necessary and is implemented specifically by each rule.
\paragraph{Dempster-Shafer rule.}
The historical rule of Dempster-Shafer~\cite{dempster2,Shafer1976} is based on the conjunctive rule, but handles the conflict as a normalization parameter.
\begin{definition}[Dempster-Shafer]\label{definition:6}
Let $m_1$ and $m_2$ be $2$ bbas over $G^\Theta$\,.
The Dempster-Shafer combination $m_1\oplusDS m_2$ of $m_1$ and $m_2$ is defined by: 
\begin{equation}\label{dmb:dsmtb2:ch2:rule:eq:2:DS}
m_1\oplusDS m_2(X)=\frac{I[X\ne\everFalse]m_1\oplusconj m_2(X)}{1-Z}\;,
\end{equation}
where $Z=m_1\oplusconj m_2(\everFalse)$ is the degree of conflict and:
$$I[X\ne\everFalse]=1\mbox{ if }X\ne\everFalse\;,\mbox{ and }I[X\ne\everFalse]=0\mbox{ if }X=\everFalse\;.$$
\end{definition}
\paragraph{Rules defined from a confict redistribution.}
Variety of rules have been proposed recently~\cite{Smarandache2005, Florea2006, Martin2006, Smarandache2006, Denoeux2006, martin2007}, especially based on a conflict redistribution~\cite{Lefevre2002}.
We will not give here a description of all these rules, but a principle of construction is proposed.
\\[5pt]
The principle is to construct a combination rule from the conjunctive rule by redistributing the conflict.
Being given a redistribution function $ r$, it is defined the combination rule $\oplus_ r$ by:
\begin{equation}\label{dmb:F2K7:ch2:rule:eq:4}
m_1\oplus[ r] m_2(X)=m_1\oplusconj m_2(X)+ r\bigl(X\big|m_1,m_2\bigr)\,m_1\oplusconj m_2(\everFalse)\,,
\end{equation}
for any $X\in G^\Theta$\,.
The redistribution function $ r$ should be typically such that:
\begin{equation}\label{eq:redistri:1}
 r(\everFalse|m_1,m_2)=-1\;,\quad\sum_{Y\in G^\Theta} r(Y|m_1,m_2)=0\;,
\end{equation}
and:
\begin{equation}\label{eq:redistri:2}
 r(X|m_1,m_2)\ge0\;,
\mbox{ for any }
X\in G^\Theta\setminus\{\everFalse\}\,.
\end{equation}
But there are examples of redistribution laws~\cite{Florea2006} where the constraint~(\ref{eq:redistri:2}) is relaxed.
However, in this approach of Florea, Jousselme and \emph{al}, the redistribution function is conditionned by the conflict $Z=m_1\oplusconj m_2(\everFalse)$, which is a more specific conditionner than $m_{1}, m_2$
\\[5pt]
There are many possible rules deduced from the redistribution principle.
Dempster-Shafer rule is among these rules:
$$
\oplusDS=\oplus[ r_{\mathrm{DS}}]
\mbox{ where }
 r_{\mathrm{DS}}(X|m_1,m_2)=\frac{m_1\oplusconj m_2(X)}{1-Z}\;,
\mbox{ for any }
X\in G^\Theta\setminus\{\everFalse\}\,.
$$
In this case, the redistribution is proportionnal to the conjunctively fused belief.
\\[5pt]
The PCR5 (Proportional Conflict Redistribution no.~5) rule~\cite{Smarandache2005,Smarandache2006} is another rule deduced from a redistribution of the conflict, \emph{i.e.}
$
\oplusPCRv=\oplus[ r_{\mathrm{PCR5}}]
$
where:
$$\begin{array}{@{}l@{}}\displaystyle
 r_{\mathrm{PCR5}}(X|m_1,m_2)=\frac{1}{Z}
\left(
\sum_{Y\in G^\Theta\setminus\{X\}}\frac{m_1(X)}{m_1(X)+m_2(Y)}m_1(X)m_2(Y)
\right.
\vspace{5pt}\\\displaystyle
\rien\hspace{120pt}\left.
+
\sum_{Y\in G^\Theta\setminus\{X\}}\frac{m_2(X)}{m_2(X)+m_1(Y)}m_1(Y)m_2(X)
\right)\;.
\end{array}$$
for any $X\in G^\Theta\setminus\{\everFalse\}$\,.
Under that form, it appears clearly that PCR5 is based on a local conflict redistribution, attributed proportionnaly to the input basic belief.
\\[5pt]
The principle of conflict redistribution implies a potentially infinite variety of rules.
The question now is \emph{how to decide for a rule or another?}
This choice is certainly dependent of the structure of the fusion problem.
A theoretical criterion for such a choice is still an issue.
\\[5pt]
In the next sections, we derive a theoretically grounded ruling method, for which the conflict redistribution is implied by a new concept of independence of the sources.
The new rules are essentially computed from an entropic optimization problem.
%
\section{Entropic approach for the rule definition}
\label{DSmTb2:Dmb:Sec:3}
To begin with this new rule concept, we directly settle the concrete optimization principles of our method.
The logical justification comes later in part~\ref{logpart}.
\subsection{Independent sources and entropy}
The idea is not completely new, and the entropy maximization principle has been linked to the conjunctive rule in~\cite{dezert}.
It is noticed, that $\oplusconj$ could be derived from an entropic maximization:
$$
m_1\oplusconj m_2(X)=\sum_{Y\cap Z=X}f_o(Y,Z)\;,\mbox{ for any }X\in G^\Theta\,,
$$
where:
\begin{equation}\label{dmb:dsmtb2:ch2:indepf:eq:4}
\begin{array}{@{\;}l@{}}\displaystyle
f_o\in\arg\max_f-\sum_{Y,Z\in G^\Theta}f(Y,Z)\ln f(Y,Z)
\vspace{5pt}\\\mbox{under constraints: }
\vspace{4pt}\\\displaystyle
\rien\quad f\ge0\;,\vspace{4pt}\\ 
\displaystyle\rien\quad\sum_{Y\in G^\Theta}f(Y,Z)=m_2(Z)\;,\mbox{ for any }Z\in G^\Theta\,,
\vspace{4pt}\\
\displaystyle\rien\quad\sum_{Z\in G^\Theta}f(Y,Z)=m_1(Y)\;,\mbox{ for any }Y\in G^\Theta\,.
\end{array}
\end{equation}
Equation~(\ref{dmb:dsmtb2:ch2:indepf:eq:4}) has a particular interpretation in the paradigm of information theory:
$f_o$ is the law with maximal ignorance of the interaction between the marginal variables $Y$ and $Z$, owing to the fact that the marginal law are $m_1$ and $m_2$.
By the way, independent sources of information should indeed provide a maximal ignorance of interaction between the maginal variables, so that the maximization of entropy appears as the good way to characterize independent sources.
Since, in this case, the constraints are limited to the marginalizations, the solution to this maximization is:
$$f_o(Y,Z)=m_1(Y) m_2(Z)\;,$$
where is recognized the joint law of two independent variables.
The definition~(\ref{dmb:dsmtb2:ch2:indepf:eq:4}) works when the \emph{open world hypothesis} is done.
But when the \emph{closed world} is hypothesized, new constaints are implied and have to be added to the optimization problem.
In regards to these considerations, a generalization of the notion of independance between variables is introduced now, and will be used for the definition of a new rule of combination.
\subsection{Independance of constrained variables}
\begin{definition}[constrained independance]
Let be given the variables $X_{1:s}$ defined over the finite sets $E_{1:s}$ respectively.
Let $\rho_o$ be a probability density defined over the joint variable $(X_{1:s})$ and let $\rho_{1:s}$ be the marginals of $\rho_o$ over the variables $X_{1:s}$ respectively.
\ Let $\vec{\varphi}(\cdot)$ be the constraint defined by
$
\vec{\varphi}:\bigl(\rho(X_{1:s})\bigr)_{(X_{1:s})\in\prod_{i=1:s}E_i}\mapsto\{false,true\}^M
$
is a vectorial Boolean mapping defined on any joint probability density $\rho$.
Then, \emph{the variables $X_{1:s}$ are independant in regards to the probability density $\rho_o$ and to the constraint $\vec{\varphi}(\cdot)\ge\vec{0}$} if and only if \emph{$\rho_o$ is the solution of the following optimization:}
\begin{equation}\label{constrained:indep:2}\begin{array}{@{}l@{}}
\displaystyle
\rho_o\in\arg\max_{\rho}\sum_{(X_{1:s})\in\prod_{i=1:s}E_i}-\rho(X_{1:s})\ln\rho(X_{1:s})\;,
\\\displaystyle
\mbox{with respect to :}
\\\displaystyle
\rien\quad\rho\ge0\;,
\\\displaystyle
\rien\quad\mbox{the marginality constraints : }\sum_{(X_{1:i-1})\in \prod_{j=1:i-1}E_{j}}
\ \sum_{(X_{i+1:s})\in \prod_{j=i+1:s}E_{j}}
\rho(X_{1:s})=\rho_i(X_i)\;,
\\\displaystyle
\rien\quad\mbox{the constraint : }\vec{\varphi}(\rho)=\overrightarrow{true}\;.
\end{array}\end{equation}
\end{definition}
It is noticed that this concept is dependent of the variables set.
For example, it is not necessary true that the variables of the subsequence $X_{i_1},\cdots,X_{i_k}$ are independant in regards to the constraints, when $X_{1:s}$ are.
There is an exception however, when there is no constraint, \emph{i.e.} $\vec{\varphi}(\rho)$ is always $\overrightarrow{true}$.
In such a case, it is proven by applying Kuhn-Tucker theorem that $\rho(X_{1:s})=\prod_{i=1:s}\rho_i(X_i)$\,.
\\[4pt]
The notion of \emph{constrained independance} is now applied to the definition of a combination rule; the sets are $E_i=G^\Theta$, and the constraint may be defined by 
$\vec{\varphi}=(\varphi_{X_{1:s}})_{X_{1:s}\in G^\Theta}$\,, where:
$$\begin{array}{@{}l@{}}\displaystyle
\varphi_{X_{1:s}}(\rho)= false \mbox{ \ if }\rho(X_{1:s})>0\mbox{ and }\bigcap_{i=1:s}X_i=\everFalse\,,
\\[4pt]\displaystyle
\varphi_{X_{1:s}}(\rho)= true \mbox{ \ otherwise.}
\end{array}$$
After simplification, this construction results in the following definition of the fusion rule.
\subsection{Entropy maximizing rule for a closed world}
\begin{definition}[entropy maximizing rule]\label{definition:7}
Let $m_{1:s}$ be $s$ bbas defined on $G^\Theta$, compliant with the \emph{closed world hypothesis}.
The combination rule~$\oplusEMR$, named Entropy Maximizing Rule (EMR), is defined by:
\begin{equation}\label{dmb:dsmtb2:ch2:rule:eq:5}
\begin{array}{@{}l@{}}\displaystyle
m_1\oplusEMR\cdots\oplusEMR m_s(X)=\sum_{\bigcap_{i=1:s}Y_i=X}f_o(Y_{1:s})\;,\mbox{ for any }X\in G^\Theta\,,
\vspace{5pt}\\\mbox{where:}
\vspace{5pt}\\
\begin{array}{@{\;}l@{}}\displaystyle
f_o\in\arg\max_f-\sum_{Y_{1:s}\in G^\Theta}f(Y_{1:s})\ln f(Y_{1:s})
\vspace{5pt}\\\mbox{under constraints: }
\vspace{4pt}\\\displaystyle
\rien\quad f\ge0\;,\vspace{4pt}\\ 
\displaystyle\rien\quad\sum_{Y_{1:i-1}, Y_{i+1:s}\in G^\Theta}f(Y_{1:s})=m_i(Y_i)\;,\mbox{ for any }i=1:s\mbox{ and }Y_i\in G^\Theta\,,
\vspace{4pt}\\
\displaystyle\rien\quad \bigcap_{i=1:s}Y_i=\everFalse \Rightarrow f(Y_{1:s})=0\;,\mbox{ for any }Y_{1:s}\in G^\Theta\,.
\end{array}
\end{array}
\end{equation}
\end{definition}
%
\begin{coroldef}\label{definition:8}
\mbox{$\oplusEMR$} is compliant with the closed world hypothesis.
\end{coroldef}
%
\paragraph{Conflict management.}
In the definition of rule $\oplusEMR$, the independence of the sources has been implemented practically by an entropic maximization principle.
From this principle, the confict management is implied automatically.
Nevertheless, the rule $\oplusEMR$ is also able to reject the sources, when they are considered as definitively non compatible.
The fundamental incompatibility of the sources is related to the feasibility of the constraints in~(\ref{dmb:dsmtb2:ch2:rule:eq:5}).
\paragraph{A glimpse of the logical background.}
The definition of $m_1\oplusEMR\cdots\oplusEMR m_s$ based on $f_o$ is the expression of hypotheses on the logical background, which require some caution when applying the rule. 
Especially, it is the joint result of a conservation of the beliefs during the combination and of a logical independence hypothesis.
The constraints in~(\ref{dmb:dsmtb2:ch2:rule:eq:5}) are also the consequence of the logical background.
In particular, the constaint $\bigcap_{i=1:s}Y_i=\everFalse \Rightarrow f(Y_{1:s})=0$ is related to the hypothesis that the sources are in coherence with the objective world.
In the most general case, the Belief functions may be used for fusing subjective information, and different logical background and constraints should be considered for such information.
However, the entropy maximization principle may be combined with many kinds of constraints, and we have potentially a general principle for defining rules. 
\section{Properties and implementation}
\label{DSmTb2:Dmb:Sec:4}
This section is devoted to the development of basic properties of EMR and to practical implementation on examples.
The following definitions and notations are used in this section.
\begin{definition}[probabilistic bba]\label{definition:9}
A bba $m$ is said to be probabilistic if there are propositions $X_1,\cdots,X_n\in G^\Theta\setminus\{\everFalse\}$ such that:
$$
\sum_{j=1}^n m(X_j)=1\mbox{ \ and \ }
\left\{\begin{array}{l@{}}\displaystyle
i\ne j\Rightarrow X_i\cap X_j=\everFalse\;,
\vspace{4pt}\\\displaystyle
\bigcup_{j=1}^n X_j=\everTrue\;.
\end{array}\right.$$
Moreover, when $X_1,\cdots, X_m$ are minimal non empty elements, that is:
$$
Y\subsetneq X_i\mbox{ implies }Y=\everFalse\mbox{ \ for any }i=1:n\;, 
$$
then $m$ is called a probability density over $G^\Theta$ and the credibility $\bel$ related to $m$ is called the probability measure on $G^\Theta$ derived from $m$. 
\end{definition}
\paragraph{Notations.}
Subsequently, $m_1, \cdots,m_s$ are bbas on $G^\Theta$ compliant with the closed world hypothesis.
It is denoted by $\bel_i\oplusEMR\bel_j$
the credibility
function related to $m_i\oplusEMR m_j$\,.
It is denoted by $\rho$ a probability density over $G^\Theta$ and by $P$ the probability measure on $G^\Theta$ derived from $\rho$\,.
\\[5pt]
Now, we have first to explain some notions about belief sharpening and its converse, belief weakening.
Belief weakening is commonly used by the community, especially when a reliability degree of the information is taken into account.
These notions are introduced now, with some refinements.
They are necessary subsequently.
\subsection{Belief sharpening}
\begin{definition}[belief sharpening]\label{definition:10}
A \emph{sharpening from $m_1$ to $m_2$} is a mapping $ r_{12}:G^\Theta\times G^\Theta\longrightarrow\Rset\;,$ such that $ r_{12}\ge0\,,$ and:
\begin{itemize}
\item $\displaystyle\sum_{Y\in G^\Theta} r_{12}(X,Y)=\sum_{Y:Y\subseteq X} r_{12}(X,Y)=m_1(X)\,,$ for any $X\in G^\Theta\,,$
\item $\displaystyle\sum_{X\in G^\Theta} r_{12}(X,Y)=\sum_{X:Y\subseteq X} r_{12}(X,Y)=m_2(Y)\,,$ for any $Y\in G^\Theta\,.$
\end{itemize}
\end{definition}
It is noticed that this definition implies:
\begin{equation}
\label{dmb:EMR:mathbbT:eq:1}
 r_{12}(X,Y) = 0\quad\mbox{for any}\quad(X,Y)\in(G^\Theta\times G^\Theta)
\mbox{ such that } Y\not\subseteq X\;.
\end{equation}
A sharpening will move the basic belief from propositions to subpropositions, thus resulting in a more accurate assignment of the basic belief.
We will also say that $m_2$ is a sharpening of $m_1$, or conversely that $m_1$ is a \emph{weakening} of $m_2$\,, when exists a sharpening from $m_1$ to $m_2$\,.
We will equivalently write the $\lhd$ relation for credibility functions:
$$
\bel_1\lhd\bel_2\stackrel{\Delta}{\iff}m_1\lhd m_2 \;.
$$
It is noticed that the sharpening implies an order relation on belief functions.
\begin{property}[equivalence relation]\label{property:1}
Define the relation $\lhd$ on belief functions by:
$$
\bel_1\lhd\bel_2
\mbox{ \ if and only if \ there is a sharpening } r_{12}\mbox{ from }\bel_1\mbox{ to }\bel_2\;.
$$
Then, the relation $\lhd$ is an order relation, and in particular:
\begin{description}
\item[$\lhd$ is transitive.] If $ r_{12}$ and $ r_{23}$ are sharpenings respectively from $\bel_1$ to $\bel_2$, and from $\bel_2$ to $\bel_3$\,, then $ r_{13}$ defined by:\footnote{Since $ r_{12}(X_1,X_2)\le m_2(X_2)$ and $ r_{23}(X_2,X_3)\le m_2(X_2)$\,, it is noticed that~(\ref{dmb:EMR:mathbbT:eq:1:2}) is well defined by setting $\frac{ r_{12}(X_1,X_2) r_{23}(X_2,X_3)}{m_2(X_2)}\stackrel{\Delta}{=}0$ when $m_2(X_2)=0$\,.}
\begin{equation}
\label{dmb:EMR:mathbbT:eq:1:2}
 r_{13}(X_1,X_3)\stackrel{\Delta}{=}\sum_{X_2\in G^\Theta}\frac{ r_{12}(X_1,X_2) r_{23}(X_2,X_3)}{m_2(X_2)}\;,
\end{equation}
is a sharpening from $\bel_1$ to $\bel_3$\,,
\item[$\lhd$ is reflexive.]More precisely, there is only one sharpening $ r$ from $\bel$ to $\bel$\,, which is defined by $ r(X,X)=m(X)$ and $ r(X,Y)=0$ for any $X,Y\in G^\Theta$ such that $Y\ne X$\,,
\item[$\lhd$ is antisymmetric.]
If $ r_{12}$ and $ r_{21}$ are sharpenings respectively from $\bel_1$ to $\bel_2$, and from $\bel_2$ to $\bel_1$\,, then: $$\rien\hspace{-20pt} r_{12}(X,X)= r_{21}(X,X)=m_1(X)=m_2(X)
\mbox{ \ and \ }
 r_{12}(X,Y)= r_{21}(X,Y)=0\;,$$
for any $X,Y\in G^\Theta$ such that $X\ne Y$\,.
\end{description}
\end{property}
Proof is done in appendix~\ref{appendix:proof:1}.
%
\begin{property}\label{property:2}
$\bel_1\lhd\bel_2$ implies $\bel_1\le\bel_2$\,.
\end{property}
Proof is done in appendix~\ref{appendix:proof:1}.
\emph{The converse is false.}

\begin{example}[counterexample]\label{example:2:1}
Assume: $$G^\Theta=\{\everFalse, A, B, C , A\cup B, B\cup C, C\cup A,\everFalse\}\;,$$ and define:
$$
m_1(A\cup B)=m_1(B\cup C)=0.5=m_2(B)=m_2(\everTrue)\;.
$$
It is obvious that $\bel_1\le\bel_2$, but it is also clear that $\bel_1\notsharp\;\bel_2$\,.
\end{example}
%
\begin{property}\label{property:3}
Probability measures are minimal for $\lhd$.
More precisely:
$$
P\,\lhd\,\bel\mbox{ \ implies \ }P=\bel\;.
$$
\end{property}
The result is a direct consequence of the definitions.

\subsection{Properties of EMR}
\label{DSmTb2:Dmb:Sec:4:subsec:prop}
%
%
%
\begin{property}[commutativity]\label{property:4}
Let $\sigma:[\![1,s]\!]\rightarrow[\![1,s]\!]$ be a permutation on $[\![1,s]\!]$.
Then:
$$
m_1\oplusEMR \cdots \oplusEMR m_s
\mbox{ \ exists if and only if \ }
m_{\sigma(1)}\oplusEMR \cdots \oplusEMR m_{\sigma(s)}
\mbox{ \ exists}\;,$$
and:
$$
m_1\oplusEMR \cdots \oplusEMR m_s
=
m_{\sigma(1)}\oplusEMR \cdots \oplusEMR m_{\sigma(s)}
\;.$$
\end{property}
The result is a direct consequence of the definitions.
%
\begin{property}[neutral element]\label{property:5}
Define the bba of total ignorance $\nu$ by $\nu(\everTrue)=1$\,.
Then:
$$
m_1\oplusEMR \cdots \oplusEMR m_s\oplusEMR\nu
\mbox{ \ exists if and only if \ }
m_1\oplusEMR \cdots \oplusEMR m_s
\mbox{ \ exists}\;,$$
and:
$$m_1\oplusEMR \cdots \oplusEMR m_s\oplusEMR\nu=m_1\oplusEMR \cdots \oplusEMR m_s\,.$$
\end{property}
Proof is given in appendix~\ref{appendix:proof:1}.
%
\begin{property}[conservation of belief]\label{property:6}
Assume that $m_1\oplusEMR\cdots \oplusEMR m_s$ exists.
Then:
\begin{equation}\label{dmb:dsmtb2:ch2:Implement:eq:1}
\bel_i\;\lhd\;\bel_1\oplusEMR\cdots\oplusEMR\bel_s\mbox{ \ for any }i=1:s
\,.
\end{equation}
\end{property}
Proof is given in appendix~\ref{appendix:proof:1}.
\begin{corollary}\label{corollary:7}
$$\bel_1\oplusEMR\cdots\oplusEMR\bel_s\ge\max_{i=1:s}\bel_i\;.$$
\end{corollary}
\begin{corollary}\label{corollary:8}
Let $X_1,\cdots,X_n\in G^\Theta$ be such that $X_i\cap X_j=\everFalse$ for any $i\ne j$\,.
Then: $$\mbox{The existence of }m_1\oplusEMR\cdots\oplusEMR m_s
\mbox{ implies}\quad
\sum_{j=1}^n\max_{i=1:s}\bel_i(X_j)\le 1 \;.$$
\end{corollary}
This result is a sufficient criterion for proving the non-existence of $m_1\oplusEMR\cdots\oplusEMR m_s$.
%
%
\begin{property}[combination of dominated belief functions]\label{property:9}
\label{Comb:dom:belief:function}
Assume that $\bel_i\,\lhd\,\bel$ for any $i=1:s$\,.
Then $\bel_1\oplusEMR\cdots\oplusEMR\bel_s$ exists.
\end{property}
\begin{proof}
Let $ r_i$ be a sharpening from $\bel_i$ to $\bel$, for $i=1:s$.
Since $ r_i(X_i,X)\le m(X)$, it is set: $$m(X)^{1-s}\prod_{i=1:s} r_i(X_i,X)\stackrel{\Delta}{=}0\quad
\mbox{when}\quad
m(X)=0\,.$$
Then, define:
$$
f(X_{1:s})=\sum_{X\in G^\Theta}m(X)^{1-s}\prod_{i=1:s} r_i(X_i,X)\;.
$$
From the definition of $ r_i(X_i,X)$, it is clear that:
$$
\prod_{i=1:s} r_i(X_i,X)=0\quad\mbox{when}\quad X\not\subset\bigcap_{i=1:s}X_i\;.
$$
As a consequence of this and of $m(\everFalse)=0$, it comes that:
$$
m(X)^{1-s}\prod_{i=1:s} r_i(X_i,X)=0\quad\mbox{when}\quad \bigcap_{i=1:s}X_i=\everFalse\;.$$ 
Then:
$$
\bigcap_{i=1:s}X_i=\everFalse
\quad\mbox{implies}\quad
f(X_{1:s})=0\,.
$$
Now, it is also derived:
$$\begin{array}{@{}l@{}}\displaystyle
\sum_{X_{1:i-1}, X_{i+1:s}}f(X_{1:s})=\sum_{X\in G^\Theta} r_i(X_i,X)m(X)^{1-s}\prod_{j=1:i-1,i+1:s}\sum_{X_{j}\in G^\Theta} r_j(X_j,X)
\vspace{5pt}\\\displaystyle
\rien\qquad\qquad=\sum_{X\in G^\Theta} r_i(X_i,X)m(X)^{1-s}m(X)^{s-1}=\sum_{X\in G^\Theta} r_i(X_i,X)=m_i(X_i)\;.
\end{array}$$
Since it is also clear that $f\ge0$\,, so there is a solution to the optimization~(\ref{dmb:dsmtb2:ch2:rule:eq:5}).
\end{proof}
%
\begin{property}[probabilistic bound]\label{property:10}
\label{section:probabilistic:bound}
EMR is coherent with the probabilistic bounds:
%
\begin{equation}
\label{dmb:F2K7:ch2:rule:eq:5-dx}
\bel_i\lhd P, \mbox{ for any i=1:s,}
\mbox{ \ implies \ }
\left\{\begin{array}{l@{}}\displaystyle
\bel_1\oplusEMR\cdots\oplusEMR\bel_s\oplusEMR P\mbox{ exists}\,,
\vspace{5pt}\\\displaystyle
\bel_1\oplusEMR\cdots\oplusEMR\bel_s\oplusEMR P = P
\;.
\end{array}\right.
\end{equation}
\end{property}
Proof is given in appendix~\ref{appendix:proof:1}.
\\[5pt]
The interpretation of this result may not be obvious.
It says that, when the sources of information are weakenings of a probabilistic law, then their fusion with this law exists and confirms the law.
However, the fused bba $\bel_1\oplusEMR\cdots\oplusEMR\bel_s$ obtained from the sources $\bel_{1:s}$ alone is not necessary compatible with a probability $P$ such that $\bel_i\lhd P$ for any $i=1:s$\,.
In particular:
$$
\left\{\begin{array}{l@{}}\displaystyle
\bel_i\lhd P, \mbox{ for any i=1:s,}
\mbox{ \ does not imply \ }
\bel_1\oplusEMR\cdots\oplusEMR\bel_s \,\lhd\, P\,,
\vspace{5pt}\\\displaystyle
\bel_i\lhd P, \mbox{ for any i=1:s,}
\mbox{ \ does not imply \ }
\left(\bel_1\oplusEMR\cdots\oplusEMR\bel_s\right)\oplusEMR P\mbox{ exists}\,.
\end{array}\right.
$$
This result is an indirect consequence of the fact that $\oplusEMR$ will add information by means of the independence principle.
This is illustrated by the following counterexample.
\begin{example}[counterexample]\label{example:11}
Let $m_1$, $m_2$ and $\rho$ be defined of over $G^\Theta$:
$$G^\Theta=\{\everFalse, A, B, C , A\cup B, B\cup C, C\cup A,\everFalse\}\;,$$
by:
$m_1(A\cup B)=m_2(A\cup B)=\rho(A)=m_1(B\cup C)=m_2(B\cup C)=\rho(C)=0.5
\,.$\\
It comes out that $\bel_1\lhd P$ and $\bel_2\lhd P$ by redistributing the masses from $A\cup B$ to $A$ and from $B\cup C$ to $C$.
Since the combination does not generate any conflict, then $m_1\oplusEMR m_2$ coincides with $m_1\oplusconj m_2$ so that:
$$
m_1\oplusEMR m_2(A\cup B)=m_1\oplusEMR m_2(B\cup C)=0.25
\mbox{ \ and \ }
m_1\oplusEMR m_2(B)=0.5\;.
$$
Obviously, $\bel_1\oplusEMR \bel_2\notsharp P$ and $(\bel_1\oplusEMR \bel_2)\oplusEMR  P$ does not exists, since otherwise $(m_1\oplusEMR m_2)\oplusEMR \rho(X)\ge0.5$ for $X=A,B,C$.
\end{example}
%
%
\begin{property}[associativity]\label{property:12}
EMR is not associative.
\end{property}
The counterexample of section~\ref{section:probabilistic:bound} turns out to be a counterexample for the associativity.
\\[5pt]
As a consequence, one have to distinguish between $(m_1\oplusEMR m_2)\oplusEMR m_3$,  $m_1\oplusEMR (m_2\oplusEMR m_3)$ and $m_1\oplusEMR m_2\oplusEMR m_3$, this last one being a simultaneous combination of $m_{1:3}$\,.
%
%
\begin{property}[idempotence]\label{property:13}
EMR is not indempotent, but it is proved:
\begin{equation}
\label{dmb:F2K7:ch2:rule:eq:5}
m\emph{ is probabilistic}\Rightarrow m\oplusEMR \cdots \oplusEMR m=m\;.
\end{equation}
\end{property}
Proof is given in appendix~\ref{appendix:proof:1}.
%
%
\subsection{Algorithm}
\label{DSmTb2:Dmb:Sec:4:subsec:algo}
The optimization~(\ref{dmb:dsmtb2:ch2:rule:eq:5}) is convex and is not difficult.
However, its size increases exponentially.
The gradient of $H(f)=\sum_{X_{1:s}}-f(X_{1:s})\ln f(X_{1:s})$ is:
$$
D_fH(f)=\sum_{X_{1:s}}-(1+\ln f(X_{1:s}))\,df(X_{1:s})\,.
$$
The following algorithm, based on the projected gradient, has been implemented:
\begin{enumerate}
\item Initialize $t=0$ and the convergence step $\theta_0$,
\item Find a feasible solution $f_0$ to the constraints of~(\ref{dmb:dsmtb2:ch2:rule:eq:5}); a simplex is solved.
If such a solution does not exist, then stop: \emph{the combination is not possible.}
Otherwise, continue on next step.
\item\label{DSmTb2:Dmb:Sec:4:subsec:algo:process:step1} Build $\Delta f_t$ by solving the quadratic program:
$$
\begin{array}{@{\;}l@{}}\displaystyle
\rien\hspace{-20pt}
\min_{\Delta f_t}\sum_{X_{1:s}}(\theta_t(1+\ln f_t(X_{1:s}))+\Delta f_t(X_{1:s}))^2\;,
\vspace{5pt}\\\mbox{under constraints: }
\vspace{4pt}\\\displaystyle
\rien\hspace{-20pt}
f_t+\Delta f_t\ge0\;,\quad
\sum_{X_{1:i-1}X_{i+1:s}}\Delta f_t(X_{1:s})=0\;,\mbox{ \ for any }i=1:s\mbox{ and any }X_{i}
\vspace{2pt}\\\displaystyle
\rien\hspace{-20pt}
\mbox{and}\quad
\bigcap_{i=1}^sX_i=\everFalse
\Rightarrow \Delta f_t(X_{1:s})=0\mbox{ \ for any }X_{1:s}\,.
\end{array}
$$
\item If $H(f_t+\Delta f_t)<H(f_t)$, then set $\theta_{t+1}=\theta_t/2$ and $f_{t+1}=f_t$\,, 
\item Else set $f_{t+1}=f_t+\Delta f_t$,
\item Set $t:=t+1$ and reiterate from step~\ref{DSmTb2:Dmb:Sec:4:subsec:algo:process:step1} until full convergence.
\end{enumerate}
%
\subsection{Experimentation}
\label{DSmTb2:Dmb:Sec:4:subsec:example}
We will consider the fusion of $m_1$ and $m_2$ defined on the power set of $\{a,b,c\}$ by:
$$\left\{\begin{array}{@{\,}l@{}}
m_1(a)=\alpha_1\,,\ 
m_1(c)=\gamma_1\,,\ 
m_1(\{a,b,c\})=1-\alpha_1-\gamma_1\;,
\vspace{4pt}\\
m_2(b)=\beta_2\,,\ 
m_2(c)=\gamma_2\,,\ 
m_2(\{a,b,c\})=1-\beta_2-\gamma_2\;.
\end{array}\right.$$
Notice that \emph{Zadeh's example} is a subcase where defined by $\alpha_1=\beta_2=0,99$ and $\gamma_1=\gamma_2=0.01$\,.
\\[5pt]
The table~\ref{F2K7:EMR:tab1} summarize the results of combnation by $\oplusEMR$ for various cases.
\begin{table*}
\caption{Example of EMR implementation.\vspace{-2pt}}
\label{F2K7:EMR:tab1}
$$
\begin{array}{@{}c|c||c|c||l@{}}\rien\vspace{-20pt}\\
\alpha_1 & \gamma_1 & \beta_2 & \gamma_2 & m=m_1\oplus m_2
\\\hline
0.99&0.01&0.99&0.01& \mathrm{Rejection} [Zadeh]
\\\hline
0.501&0&0.501&0& \mathrm{Rejection}
\\\hline
0.499&0&0.499&0& m(a)=m(b)=0.499\,,\ m(\{a,b,c\})=0.002
\\\hline
 0.3&0.1
&0.3&0.1
& m(a)=m(b)=0.3\,,\ m(c)=0.175\,,\ m(\{a,b,c\})=0.225
\\\hline
 0.3&0.05
&0.3&0.05
& m(a)=m(b)=0.3\,,\ m(c)=0.09375\,,\ m(\{a,b,c\})=0.30625
\\\hline
 0.3&0.01
&0.3&0.01
& m(a)=m(b)=0.3\,,\ m(c)=0.01975\,,\ m(\{a,b,c\})=0.38025
\end{array}\vspace{-5pt}$$
\end{table*}
These examples express clearly the property of belief conservation, and the capacity of EMR to reject any combination which is not compatible with this property.
\section{Conclusion}
The rule EMR is a theoretically grounded combination rule, defined on the basis of:
\begin{itemize}
\item constraints expressing the belief conservation, which is a desirable property,
\item an hypothesis of sources independence, which is generalized by means of a principle of information maximization.
\end{itemize}
The good news is that such a rule exists.
Especially, it is complient with the probabilistic bound property as expressed in section~\ref{section:probabilistic:bound}.
The bad news is that such a rule is quite constraining and cannot be associative.
Here comes a fundamental point related to the use of the hypothesis of independence.
This hypothesis introduces information which may imply some conflicts with the \emph{ground truth} expressed by means of a probability of reference $P$.
The properties of sections~\ref{Comb:dom:belief:function} (combination of dominated belief functions) and~\ref{section:probabilistic:bound} express that.
More precisely, it is possible to combine as many sources of information (even $P$ itself!) which is compatible with $P$, so as to make the best approximation of $P$:
$$
\bel_i\lhd P, \mbox{ for any i=1:s,}
\mbox{ \ implies \ }
\bel_1\oplusEMR\cdots\oplusEMR\bel_s\mbox{ exists}
\;.
$$ 
But as soon as the rule has been applied, the introduction of information by means of the independence principle may implies conflicts with the \emph{ground truth}:
$$
\bel_i\lhd P, \mbox{ for any i=1:s,}
\mbox{ \ does not imply \ }
\bel_1\oplusEMR\cdots\oplusEMR\bel_s \,\lhd\, P\,.
$$
This finding leads us to question about the hypotheses of implementation of belief combination rules, at least in the context of a closed universe.
We could give up with the property of belief conservation: this is actually what most rules do in practice, when redistributing the conjunctive conflict.
We could resign ourself to implement the combination rule just as a final step of the fusion process, as this is suggested by the properties of EMR.
However, such approach is practically impossible in regards to the combinatorics.
But there is perhaps a third way: we could manipulate conditionnal propositions instead of absolute propositions, as is usually done: conditional propositions are efficient tools for handling the independence hypotheses.
Conditionnal beliefs are not a new concept; the theoretical tools exist.
Such an approach would lead to define some forms of reasoning similar to the Bayesian reasoning.
\part{A multimodal logic for modelling beliefs}
\label{logpart} 
Part~\ref{beliefpart} introduced the rule EMR by applying a principle of sources independence based on a maximization of the information.
While the resulting optimization~(\ref{dmb:dsmtb2:ch2:rule:eq:5}) appears as a natural extension in regards to the conjunctive rule, this problem and especially its constraints have not been justified clearly.
The prupose of this second part is to address the question of belief combination from a logical point of view.
Based on this logical interpretation, the definition~(\ref{dmb:dsmtb2:ch2:rule:eq:5}) is implied.
\section{Belief functions, modality and probability}
Belief functions may be seen as a generalization of probabilistic distributions by combining imprecisions within the uncertainty.
Belief functions are caracterized by the notion of basic belief assignment which will assign a basic belief to any element of the Boolean algebra.
In this way, bbas may be seen as a generalization of probabilistic densities, which assign probabilistic densities to the minimal elements of the Boolean algebra.
This difference implies the ability of belief functions for manipulating imprecise information.
It has also an algebraic consequence, by replacing the additivity property by the \emph{sub-additivity} property.
The qualification of belief functions as a generalization of probabilities has been questionned, by arguing that belief functions were subcases of random sets.
While this is true in some manner, an elementary interpretation of this theory by means of random sets is not able to seize some semantic subtilities behind belief functions.
\\[5pt]
However, there is another way to concile probabilities and belief functions.
This could be done by introducing the semantic at a logical level.
While probabilities are constructed on a classical logical framework (Boolean algebras), it is still possible to implement probabilities on a non classical logic, and this how we will interpret belief functions.
In~\cite{ruspini1987}, Ruspini derived a logical counterpart of the credibility function by means of an epistemic operator $K$ representing the knowledge of the source of information.
This operator was in fact related to the modal operator of system S5.
He proposed a logical counterpart of basic belief assignment, \emph{i.e.} the \emph{epistemic} proposition denoted $e(X)$, derived from logical combinations of the modal propositions as follows:
\begin{equation}\label{eq:ruspini:1}
e(X)=KX\setminus\bigcup_{Y\subset X; Y\ne X}KY\;.
\end{equation}
Subsequently, an equivalent construction is proposed on the basis of a multimodal logic.
In our model, $\bel_i(X)$, the credibility of $X$ according to sensor $i$, is interpreted as the probability $P([i]X\wedge\neg[i]\bot)$ of the modal proposition $[i]X\wedge\neg[i]\bot$\,.
\\[5pt]
Ruspini also introduced the principle of a logical combination rule: 
\begin{equation}\label{eq:ruspini:2}
e(X)=\bigcup_{Y\cap Z=X}[e_1(Y)\cap e_2(Z)]\;.
\end{equation}
This combination rule is in fact an immediate consequence of an \emph{elaborated} axiom directly inspired from the conjunctive combination rule formulated on the credibility functions:
\begin{equation}\label{eq:ruspini:3}
KX\mbox{ is true }
\iff
\mbox{ there is } Y,Z \mbox{ such that } Y\cap Z\subseteq X \mbox{ and } K_1Y, K_2Z \mbox{ are true.}
\end{equation}
Subsequently, we will go a step further in this way by deriving the logical combination rule~(\ref{eq:ruspini:2}) from axioms of a multimodal logic.
These axioms will be related to elementary properties of our deduction system.
\\[5pt]
From~(\ref{eq:ruspini:2}), Ruspini derived back Dempster-Shafer combination by applying the property $P(e_1(Y)\cap e_2(Z))=P(e_1(Y))P(e_2(Z))$ \,, which depicts a rough independence hypothesis of the sources.
Since $e_1(Y)\cap e_2(Z)=\everFalse$ whenever $Y\cap Z=\everFalse$, this approach generates conflict and needs a normalization of the combination rule.
In particular, it is possible to have $P(\everFalse)>0$\,;
this is clearly not satisfying from a logical viewpoint.
\\[5pt]
Again, we will go further with this idea of Ruspini, but, this time, by implementing our principle of sources independence based on the information maximization.
This principle takes care of the logical exclusions.
Nevertheless, we will also handle the distinction between the open world and the closed world, by considering different axioms.
Especially, closed worlds will be such that the modalities of the contradiction are necessarily a contradiction, \emph{i.e.} $[i]\bot\equiv\bot$\,, while open worlds will accept a relaxation of this constraint.
%
\section{Definition of a multimodal logic}
\label{logpart:sect:1} 
This section introduces a multimodal logic which is the backbone for a logical interpretaion of the rule EMR.
This logic has been previously introduced in~\cite{DSmTb1:dmb}.
%
%
%
\subsection{Language definition}
\label{DSmTb2:Dmb:Sec:5:subsec:lang:def}
It is defined a \emph{finite} 
set, $\mathbf{P}$, called the set of atomic propositions.
Atomic propositions of $\mathbf{P}$ are denoted $p,q,r,\cdots$
For simplicity, it will be assumed subsequently that \emph{the atomic propositions describe a partition}, that is the conjunction of different atomic propositions is the contradiction and the disjunction of all atomic propositions is a tautology.
Although these strong hypotheses depart from usual axiomatization of classical logic, it is not a restriction, since it is known that is possible to construct a free Boolean algebra by means of a power set.
\\[3pt]
It is defined the finite set $I=\{1,\cdots,n\}$, describing $n$ sources of information.
\begin{definition}[classical propositions]\label{definition:11}
The set~$\mathcal{L}_C$ of classical propositions is defined recursively by $\bot\in\mathcal{L}_C$\,, $\mathbf{P}\subseteq\mathcal{L}_C$ and $X\rightarrow Y \in \mathcal{L}_C$ for any $X,Y\in\mathcal{L}_C$\,.
\end{definition}
The classical propositions describe the logical information on the real world.
\begin{definition}[simple modal propositions]\label{definition:12}
The set~$\simplemodal$ of simple modal propositions are defined recursively by $\bot\in\simplemodal$\,, $\mathbf{P}\subseteq\simplemodal$\,,
$X\rightarrow Y \in \simplemodal$ for any $X,Y\in\simplemodal$\,,
and $[J]X \in \simplemodal$ for any $X\in\mathcal{L}_C$ and any \emph{non empty} $J\subseteq I$\,.
\end{definition}
In $\simplemodal$, modalities are only applied to classical propositions, resulting in a language free of nested modalities.
This language contains all necessary elements for defining the belief reasonning at a logical level.
Nevertheless, it is not closed under the modal operators, and the language of all modal propositions is prefered for this logical study.  
\begin{definition}[modal propositions]\label{definition:13}
The set~$\mathcal{L}$ of modal propositions is defined recursively by $\bot\in\mathcal{L}$\,, $\mathbf{P}\subseteq\mathcal{L}$\,,
$X\rightarrow Y \in \mathcal{L}$ for any $X,Y\in\mathcal{L}$\,,
and $[J]X \in \mathcal{L}$ for any $X\in\mathcal{L}$ and any \emph{non empty} $J\subseteq I$\,.
\end{definition}
Propositions of $\mathcal{L}$ are denoted $X,Y,Z,\cdots$
It is noticed that the \emph{empty} modality $[\emptyset]X$ is not defined in $\mathcal{L}$\,.
The modal propositions describe the logical information according to the sources of information.
Subsequently, the modal operator $[J]$ describes logically the information received and interpreted by the set $J$ of sources of information working conjointly.
Practically, we are essentially interested by the information on the classical propositions, so that, although we authorize nested modal propositions like $[J][K]X$, we will not deal with them in practice.
\\[5pt]
The abbreviations
\mbox{$\neg X\stackrel{\Delta}{=}X\rightarrow\bot\,,$}
\mbox{$X\vee Y\stackrel{\Delta}{=}\neg X\rightarrow Y\,,$}
\mbox{$X\wedge Y\stackrel{\Delta}{=}\neg (X\rightarrow \neg Y)\,,$}
\mbox{$\top\stackrel{\Delta}{=}\neg\bot$}
are used, for any $J\subseteq I$ and $X,Y\in\mathcal{L}$\,.
The dual of $[J] X$ is defined as \mbox{$<\!J\!>X \stackrel{\Delta}{=}\neg [J]\neg X\,.$}
\\[3pt]
A multimodal logic is defined subsequently, by means of axioms and deductions rules.
A proposition which is deduced from these axioms and rules is called a \emph{theorem}. 
The following definition will be usefull subsequently:
\begin{definition}\label{definition:14}
For any \emph{finite} set of propositions $E\subset\mathcal{L}$, it is defined:
\begin{equation}\label{vee:E:def}
\setvee E=\bigvee_{X\in E}X\;.
\end{equation}
is often used subsequently.
\end{definition}
\begin{definition}[logical equivalence]\label{definition:15}
The logical equivalence $\equiv$ is defined by:
$$
X\equiv Y
\mbox{ if and only if }
X\rightarrow Y\mbox{ and }Y\rightarrow X\mbox{ are theorems.}
$$
\end{definition}
\subsection{Axioms.}
\label{DSmTb2:Dmb:Sec:5:subsec:lang:axioms}
The multimodal logic is defined by the following rules and axioms defined for any propositions $X,Y\in\mathcal{L}$ and subsets $J,K\subseteq I$\,:
\begin{description}
\item[Classical axioms.]
\item[Atomic propositions form a partition.]
$\setvee{\mathbf{P}}$ is an axiom and
$(p\wedge q)\rightarrow\bot$ is an axiom\,, for any atomic propositions $p,q\in\mathbf{P}$ such that $p\ne q$\,,
\item[Modus ponens.]If $X$ and $X\rightarrow Y$ are theorems, then $Y$ is a theorem\,,
\item[Modal rules and axioms:]
\item[\rien$\quad N$.] If $X$ is a theorem, then $[J]Y$ is a theorem\,,
\item[\rien$\quad K$.] $[J](X\rightarrow Y)\rightarrow([J]X\rightarrow [J]Y)$ is an axiom\,,
\item[\rien$\quad T$.] $[J]X\rightarrow X$ is an axiom \ (optional),
\item[\rien$\quad Inc$.] $[J]X\rightarrow [J\cup K] X$ is an axiom\,,
\item[\rien$\quad Ind$.] Let $E\subseteq\mathbf{P}$ be a subset of atomic propositions.
Using the notation~(\ref{vee:E:def}), the following proposition is an axiom:
$$[J\cup K]\setvee E\rightarrow \bigvee_{\mbox{\scriptsize$\begin{array}{@{}l@{}}F,G\subseteq\mathbf{P}\\F\cap G\subseteq E\end{array}$}}\left(\left([J] \setvee F\right)\wedge \left([K]\setvee G\right)\right)\;.
$$
\end{description}
The hypothesis, that \emph{atomic propositions form a partiotion}, is used for simplicity. On the first hand, it allows the introduction of general Boolean structures and not only free Boolean structures for characterizing the model of the world.
On the second hand, it makes much more simple the definition of axiom $Ind$ as well as the subsequent definition of the basic logical assignment. 
The rule $N$ and axiom $K$ are minimal requirements of modal systems, and express that the sources are logically coherent.
The axiom $T$ is optional and expresses that the sources provide information which are coherent with the ground truth.
We will see that this axiom is in relation with the closed world hypothesis. 
The axiom $Inc$ expresses that the sources are always contributing positively to the global knowledge.
Incidentally, this axiom also expresses that the sources are working in coherence with each other at a the logical level.
The axiom $Ind$ expresses at the logical level that the sources are contributing \emph{independently} to the global knowledge of the classical propostions, that is the joint knowledge $[J\cup K]X$ does not contains more information than the disjoint knowledges $[J] Y\wedge [K]Z$, where $X,Y,Z$ are classial propositions such that $(Y\cap Z)\rightarrow X$\,.
\\[5pt]
It is assumed subsequently that the reader is familiar with classical propositional logic and its main results.
Some results are recalled subsequently about the logical equivalence.
\subsection{Logical equivalence}
Owing to the axioms and rules of classical logic, $\equiv$ is an equivalence relation on the propositions of $\mathcal{L}$.
The equivalence class of a proposition $X\in\mathcal{L}$ for the logical equivalence is denoted $\equivclass{X}$.
Let $\mathcal{F}\subseteq\mathcal{L}$\,.
The set of equivalence classes of propositions $X\in\mathcal{F}$ is also denoted $\equivclass{F}$\,.
\\[3pt]
The logical operators are naturally infered on the classes of propositions.
For example, it is defined
$\equivclass{X}\wedge\equivclass{Y}\stackrel{\Delta}{=}\equivclass{X\wedge Y}$\,, $[J]\equivclass{X}\stackrel{\Delta}{=}\equivclass{[J]X}\,,$\dots
An order relation is also defined over $\equivclass{\mathcal{L}}$.
\begin{definition}[logical order]\label{definition:16}
$$
\equivclass{X}\subseteq\equivclass{Y}\mbox{ if and only if }X\rightarrow Y\mbox{ is a theorem.}
$$
The notation $\equivclass{X}\subsetneq\equivclass{Y}$ is used when $\equivclass{X}\subseteq\equivclass{Y}$ and $\equivclass{X}\neq\equivclass{Y}$\,.
\end{definition}
Owing to the axioms, there is an morphism between $2^{\mathbf{P}}$ and $\equivclass{\mathcal{L}_C}$.
:
\begin{definition}[morphism]\label{definition:17}
For any classical proposition $X\in\mathcal{L}_C$\,, it is defined $pX$ be the set of atomic propositions such that $\equivclass{p}\subseteq\equivclass{X}$.
\begin{equation}
\label{def:pX:from:X}
pX=\left\{p\in\mathbf{P}\;\big/\;\equivclass{p}\subseteq\equivclass{X}\right\}\;.
\end{equation}
\end{definition}
\begin{property}[morphism]\label{property:13:1}
Any classical proposition is equivalent to proposition $\setvee pX$\,:
$$
\forall X\in\mathcal{L}_C\;,\ 
\equivclass{X}= \equivclass{\setvee pX}\,.
$$
For any $E,F\subseteq \mathbf{P}$\,, it is also derived:
$$
(\setvee E)\wedge(\setvee F)\equiv\setvee(E\cap F)\,,\ (\setvee E)\vee(\setvee F)\equiv\setvee(E\cup F)\;.
\mbox{ \ and \ }\neg \setvee E\equiv\setvee(\mathbf{P}\setminus E)\;.
$$
\end{property}
These are clear consequences of classical propositional logic.
\section{Deriving the combination rules}
\label{logpart:sect:3}
%
\subsection{Preliminary results}
Some useful logical theorems are now derived.
These theorems are deduced without the help of the optional axiom $T$.
%
\begin{property}[modality is non-decreasing]\label{property:14}
\begin{equation}\label{F2K7:EMR:small:prop:1}
\mbox{If }X\rightarrow Y\mbox{ is a theorem, then }
[J]X\rightarrow [J]Y\mbox{ is a theorem.}
\;.
\end{equation}
\end{property}
This result is a direct consequence of $N$ and $K$\,.
%
%
\begin{property}[modality and conjunction]\label{property:15}
\begin{equation}\label{F2K7:EMR:small:prop:2}
([J]X\wedge[J]Y)\equiv[J](X\wedge Y)\;.
\end{equation}
\end{property}
This result is a consequence of the theorems $(X\wedge Y)\rightarrow X$\,,  $(X\wedge Y)\rightarrow Y$\,, $X\rightarrow(Y\rightarrow(X\wedge Y))$\,, $N$ and $K$.
%
%
\begin{property}[modality and disjunction]\label{property:16}
\begin{equation}\label{F2K7:EMR:small:prop:3}
([J]X\vee[J]Y)\rightarrow[J](X\vee Y)\mbox{ is a theorem.}
\end{equation}
\end{property}
This result is a consequence of the theorems $X\rightarrow (X\vee Y)$\,,  $Y\rightarrow (X\vee Y)$\,, $N$ and $K$.
%
\begin{property}[conjunction of heterogeneous modalities]\label{property:17}
\begin{equation}\label{F2K7:EMR:small:prop:4}
([J]X\wedge[K]Y)\rightarrow[J\cup K](X\wedge Y)\;.
\end{equation}
\end{property}
Deduced from theorems $[J]X\rightarrow[J\cup K]X$ and $[K]Y\rightarrow[J\cup K]Y$\,, derived from $Inc$.
%
\begin{property}[disjunction of heterogeneous modalities]\label{property:18}
\begin{equation}\label{F2K7:EMR:small:prop:5}
([J]X\vee[K]X)\rightarrow[J\cup K]X\;.
\end{equation}
\end{property}
Immediate consequence of $Inc$.
\subsection{Logical combination rule}
\label{DSmTb2:Dmb:Sec:5:subsec:intro}
\subsubsection{Preliminary definitions}
\label{DSmTb2:Dmb:Sec:5:subsec:rule:preldef}
The notion of partition will be useful for defining the logical combination.
\begin{definition}[partition]\label{definition:18}
A set of propositions, $\Pi\subseteq\mathcal{L}$ is a partition of $\top$ if it satisfies:
\begin{itemize}
\item The propositions of $\Pi$ are exclusive : $X\wedge Y\equiv\bot$ for any $X,Y\in \Pi$ such that $X\ne Y$\,,
\item The propositions of $\Pi$ are exhaustive: $\bigvee_{X\in \Pi}X\equiv\top$\,.
\end{itemize}
\end{definition}
Notice that $\Pi$ may contain $\bot$, in this definition.
The following property is a direct consequence of the definition.
\begin{property}\label{property:20}
Let $\Pi$ and $\Lambda$ be two partitions of $\top$.
Then:
\begin{equation}\label{partition:prop:1}\begin{array}{@{}l@{}}\displaystyle\bullet\quad
\left(\bigvee_{X\in A}X\right)\wedge\neg\left(\bigvee_{X\in B}X\right)
\equiv\bigvee_{X\in A\setminus B}X\;,\mbox{ \ for any } A,B\subseteq\Pi\,,
\vspace{5pt}\\\displaystyle\bullet\quad
X_1\wedge Y_1\equiv X_2\wedge Y_2\mbox{ and }(X_1, Y_1)\ne(X_2, Y_2)\mbox{ \ imply \ }X_1\wedge Y_1\equiv X_2\wedge Y_2\equiv\bot\;,
\vspace{3pt}\\\displaystyle
\rien\hspace{250pt}\mbox{ \ for any } X_1,Y_1\in\Pi\mbox{ and }X_2,Y_2\in\Lambda\;,
\vspace{5pt}\\\displaystyle\bullet\quad
\mbox{The set of joint propositions }\Gamma=\{X\wedge Y\;/\;X\in \Pi\mbox{ and }Y\in \Lambda\}\mbox{ is a partition.}
\end{array}
\end{equation}
\end{property}
%
\subsubsection{Logical combination}
\label{DSmTb2:Dmb:Sec:5:subsec:rule:preldef:2}
%
\paragraph{Principle of the logical combination.}
Let $J,K\subseteq I$ be sources subsets, and $X$ be any classical proposition.
Arises the following issue: \emph{How to characterize the fused information \mbox{$[J\cup K]X$} from the primary information \mbox{$[J]X$} and \mbox{$[K]X$}\,.}
In order to solve this question, we introduce first the notion of \emph{basic logical assignments} (bla) which constitute the elementary logical components of the information.
\begin{definition}[basic logical assignments]\label{definition:19}
Let $E\subseteq \mathbf{P}$
The basic logical assignment (bla) related to the classical proposition $\setvee E$ and to sources subset $J$ is the modal proposition $(J)\setvee E$ defined by:
\begin{equation}
\label{DSmTb2:Dmb:Sec:5:subsec:rule:eq:1}
(J)\setvee {E}\stackrel{\Delta}{=}[J]\setvee {E}\wedge\neg\left(
\bigvee_{F\subsetneq E}[J]\setvee{F}
\right)\;.
\end{equation}
\end{definition}
The bla $(J)\setvee  E$ is the logical information, which the sources of set $J$ attribute to proposition $\setvee  E$ intrinsically.
The information of $(J)\setvee  E$ cannot be attributed to \emph{classical} propositions smaller than $\setvee  E$.
From the morphism property~\ref{property:13:1}, the definition extends to $\equivclass{\mathcal{L}_C}$\,:
\begin{definition}[bla on classes]\label{definition:20}
For any $\equivclass{X}\in\equivclass{\mathcal{L}_C}$\,, the basic logical assignment for $\equivclass{X}$ is defined by $(J)\equivclass{X}\stackrel{\Delta}{=}(J)\equivclass{\setvee {pX}}\,.$
The bla $(J)\equivclass{X}$ is also alternatively defined by:
\begin{equation}
\label{DSmTb2:Dmb:Sec:5:subsec:rule:eq:1bis}
(J)\equivclass{X}=[J]\equivclass{X}\wedge\neg\left(
\bigvee_{\equivclass{Y}\in\equivclass{\mathcal{L}_C}:\equivclass{Y}\subsetneq \equivclass{X}}[J]\equivclass{Y}
\right)\;
.
\end{equation}
\end{definition}
The blas have interesting properties, which will imply the definition of a logical combination. 
\begin{property}[exclusivity]\label{property:25}
The blas $(J)\setvee E$, where $E\subseteq \mathbf{P}$, are exclusive:
\begin{equation}
\label{DSmTb2:Dmb:Sec:5:subsec:rule:eq:2}
\forall E,F\subseteq\mathbf{P}\,,\;
E\ne F\Rightarrow (J)\setvee{E}\wedge(J)\setvee{F}\equiv\bot\;.
\end{equation}
\end{property}
Proof is given in appendix~\ref{appendix:modal:proof:1}.
\begin{property}[exhaustivity]\label{property:26}
The blas $(J)\setvee E$, where $E\subseteq \mathbf{P}$, are exhaustive:
\begin{equation}
\label{DSmTb2:Dmb:Sec:5:subsec:rule:eq:3}
\bigvee_{F\subseteq E}
(J)\setvee F\equiv[J]\setvee E\;,
\mbox{ and in particular: }
\bigvee_{F\subseteq \mathbf{P}}(J)\setvee F\equiv\top\;.
\end{equation}
\end{property}
Proof is given in appendix~\ref{appendix:modal:proof:1}.
\begin{corollary}[inversion]\label{property:27}
It is possible to define $[J]\equivclass{X}$ from the blas $(J)\equivclass{Y}$\,:
\begin{equation}
\label{DSmTb2:Dmb:Sec:5:subsec:rule:eq:3:bis}
\bigvee_{
[J]\equivclass{X}=\equivclass{Y}\in\equivclass{\mathcal{L}_C}:\equivclass{Y}\subseteq\equivclass{X}}
(J)\equivclass{Y}\;,
\mbox{ \ for any }\equivclass{X}\in\equivclass{\mathcal{L}_C}\,.
\end{equation}
\end{corollary}
It is deduced from these results that bbas constitute partitions:
\begin{corollary}[bla partition]\label{property:28}
The blas $(J)\setvee E$, where $E\subseteq \mathbf{P}$, constitute a partition of $\top$\,.
\end{corollary}
\begin{corollary}[joint bla partition]\label{property:29}
Let $J,K\subseteq I$.
The propositions $(J)\setvee E\wedge(K)\setvee F$, where $E,F\subseteq \mathbf{P}$, constitute a partition of $\top$\,.
\end{corollary}
Based on these partitions, it is possible to express the logical combination:
\begin{property}[computing the combination]\label{property:30}
\begin{equation}\label{DSmTb2:Dmb:Sec:5:subsec:rule:eq:4}
(J\cup K)\setvee E\equiv\bigvee_{F\cap G=E}
\left((J)\setvee F\wedge(K)\setvee G\right)\;,
\mbox{ \ for any }E\subseteq\mathbf{P}\,.
\end{equation}
\end{property}
\begin{corollary}\label{property:31}
\begin{equation}\label{DSmTb2:Dmb:Sec:5:subsec:rule:eq:4bis}
(J\cup K)\equivclass{X}=\bigvee_{\equivclass{Y},\equivclass{Z}\in\equivclass{\mathcal{L}_C}:\equivclass{Y}\wedge \equivclass{Z}=\equivclass{X}}
\left((J)\equivclass{Y}\wedge(K)\equivclass{Z}\right)\;,
\mbox{ \ for any }\equivclass{X}\in\equivclass{\mathcal{L}_C}\,.
\end{equation}
\end{corollary}
The proof of the main property will apply the following lemma.
\begin{lemma}\label{property:32}
$$
[J\cup K]\setvee E\equiv\bigvee_{F\cap G\subseteq E}([J]\setvee F\wedge[K]\setvee G)
\equiv\bigvee_{F\cap G\subseteq E}\left((J)\setvee F\wedge(K)\setvee G\right)\;.
$$
\end{lemma}
Proofs are given in appendix~\ref{appendix:modal:proof:1}.
\begin{property}[generalized combination]\label{property:33}
The combination~(\ref{DSmTb2:Dmb:Sec:5:subsec:rule:eq:4bis}) happens to be associative and may be extended to $s$ sources:
\begin{equation}\label{DSmTb2:Dmb:Sec:5:subsec:rule:eq:4:extended}
(J_1\cup\cdots\cup J_s)\equivclass{X}=\bigvee_{\equivclass{Y_{1:s}}\in\equivclass{\mathcal{L}_C}:\bigwedge_{i=1:s}\equivclass{Y_i}=\equivclass{X}}
\bigwedge_{i=1:s}(J_i)\equivclass{Y_i}\;,\mbox{ \ for any }\equivclass{X}\in\equivclass{\mathcal{L}_C}\;.
\end{equation}
\end{property}
\subsubsection{Processus of logical fusion}
As a conclusion, the fused sources $J\cup K$ is computed from sources $J$ and $K$ as follows:
\begin{itemize}
\item Build $\displaystyle(H)\equivclass{X}$ by means of~(\ref{DSmTb2:Dmb:Sec:5:subsec:rule:eq:1bis})
 for $H=J,K$\,,
\item Compute $\displaystyle(J\cup K)\equivclass{X}$ by means of~(\ref{DSmTb2:Dmb:Sec:5:subsec:rule:eq:4bis})
\item Derive $\displaystyle[J\cup K]\equivclass{X}$ by means of~(\ref{DSmTb2:Dmb:Sec:5:subsec:rule:eq:3:bis}), 
\end{itemize}
for any $\equivclass{X}\in\equivclass{\mathcal{L}_C}$\,.
\section{Model and consequences}
\label{logpart:sect:2}
In this section, a model is constructed for the logic.
This model is not complete, when embeded modalities are considered.
However, if we restrict to the set of simple modal propositions, $\simplemodal$, the model has some completude properties, which will be useful for deriving the belief combination from the logical combination. 
The model is not based on a Kripke constuction.

\subsection{Model}
\subsubsection{Definition}
Two models are constructed for the logic respectively without and with the optional axiom $T$.
The models are quite similar, except for the minimal elements, denoted $\phi(\cdots)$.
The elements $\phi(\xModel_0,\YModel_{1:n})$ are defined for any $\xModel_0\in \mathbf{P}$ and any $\YModel_{1:n}\subseteq \mathbf{P}$ by:
\begin{itemize}
\item \emph{Without} axiom $T$\,:
\begin{equation}\label{model:def:min:elt:withoutT}
\phi(\xModel_0,\YModel_{1:n})=\{(\xModel_0,\YModel_{1:n})\}\;,
\end{equation}
\item \emph{With} axiom $T$\,:
\begin{equation}\label{model:def:min:elt:withT}
\begin{array}{@{}l@{}}\displaystyle
\phi(\xModel_0,\YModel_{1:n})=\{(\xModel_0,\YModel_{1:n})\}\;,\mbox{ if }\xModel_0\in\bigcap_{i=1}^n\YModel_i\;,
\\\displaystyle
\phi(\xModel_0,\YModel_{1:n})=\{\}\mbox{ otherwise.}
\end{array}
\end{equation}
\end{itemize}
For any proposition $p\in\mathbf{P}$\,, it is defined: $$\modelof{p}=\bigcup_{\YModel_{1:n}\subseteq\mathbf{P}}\phi(p,\YModel_{1:n})\;.$$
For any $\YModel\subseteq\mathbf{P}$\,, it is also defined:
$\modelof{E}=\bigcup_{\xModel\in\YModel}\modelof{\xModel}\,.$
\\
Then, the model $\modelalt{X}$ of any proposition $X\in\mathcal{L}$ is defined recursively by:
\begin{itemize}
\item $\modelalt{p}=\modelof{p}$ for any $p\in\mathbf{P}$ and $\modelalt{\top}=\modelof{\mathbf{P}}$\,,
\item $\modelalt{X\wedge Y}=\modelalt{X}\cap\modelalt{Y}$\,,
$\modelalt{X\vee Y}=\modelalt{X}\cup\modelalt{Y}$
and
$\modelalt{\neg X}=\modelalt{\top}\setminus\modelalt{X}$\,,
for any $X,Y\in\mathcal{L}$\,,
\item For any non empty $J\subseteq I$ and $X\in\mathcal{L}$, it is defined:
\begin{equation}
\label{def:model:modal}
\modelalt{[J]X}
=
\bigcup_{\xModel_0\in\mathbf{P}}
\bigcup_{\YModel_{I\setminus J}\subseteq\mathbf{P}}
\bigcup_{\YModel_{J}\subseteq\mathbf{P}:\modelof{\bigcap_{j\in J}\YModel_j}\subseteq\modelalt{X}}
\phi(\xModel_0,\YModel_{1:n})
\end{equation}
\end{itemize}
Other connectors are defined from $\vee$\,, $\neg$ and $[J]$\,, \emph{e.g.} $\modelalt{X\rightarrow Y}=\modelalt{\neg X\vee Y}$ and $\modelalt{<J>X}=\modelalt{\neg [J]\neg X}$\,.
\begin{property}[classical propositions]\label{property:model:1}
Let $X\in\mathcal{L}_C$ be a classical proposition.
Then, it is clear from the definition, that there is $\YModel\subset\mathbf{P}$ such that $\modelalt{X}=\modelof{\YModel}$.
\end{property}
\subsubsection{Interpretation}
Subsequently, we are assuming $X,X_i,Y\in\mathcal{L}_C$ and $p\in\mathbf{P}$\,.
It is assumed $\YModel_i\subset X_i$ such that $\modelalt{X_i}=\modelof{\YModel_i}$\,.
It is easily shown that:
\begin{equation}
\label{def:model:modal:equiv}
\modelalt{p\wedge\bigwedge_{i=1:n}(i)X_i}
=\phi(p,\YModel_{1:n})\;.
\end{equation}
In fact, it is implied from propositions~(\ref{DSmTb2:Dmb:Sec:5:subsec:rule:eq:3}) and~(\ref{DSmTb2:Dmb:Sec:5:subsec:rule:eq:4:extended}) of the previous section that any combination of non-nested modal propositions, like $[J]X$ with $X\in\mathcal{L}_C$, by means of classical operators ($\wedge,\vee,\neg,\rightarrow$) is equivalent to a disjunction of bla like $\bigwedge_{i=1:n}(i)Y$ with $Y\in\mathcal{L}_C$\,.
As a consequence, any propositions of $\simplemodal$, the set of simple modal propositions, is equivalent to a disjunction of propositions like $p\wedge\bigwedge_{i=1:n}(i)X_i$\,.
Now, in the case where axioms $T$ also holds, it is easily derived that:
$$
p\wedge\bigwedge_{i=1:n}(i)\setvee \YModel_i
\equiv\bot\;,
\mbox{ for any }\YModel_{1:n}\subseteq \mathbf{P}\mbox{ such that }p\not\in\bigcap_{i=1:n}E_i\;,
$$
which is the logical counterpart of the definition of~(\ref{model:def:min:elt:withT}).
It follows from these remarks that our model (assuming that it fulfills actually the axioms) have \emph{completude property while restricted to $\simplemodal$,} so that:
\begin{equation}
\label{def:model:modal:equal:equiv}
\equivclass{X}=\equivclass{Y}
\mbox{ if and only if }
\modelalt{X}=\modelalt{Y}\;,\mbox{ \ for any }X,Y\in \simplemodal\,.
\end{equation}
In the case, where axiom $T$ does hold, it is also noticed that the following characteristic constraints is obtained for the simple modal propositions:
\begin{equation}
\label{def:model:modal:versionof:T}
\modelalt{\bigwedge_{i=1:s}Y_i}=\modelalt{\bot}
\mbox{ implies }
\modelalt{\bigwedge_{i=1:s}(J_i)\equivclass{Y_i}}=\modelalt{\bot}\;,\mbox{ \ for any }Y_{1:s}\in \mathcal{L}_C\,.
\end{equation}
From the model definition, it is also clear that this constraint is sufficient to infer the consequence of axiom $T$ on the modal propostions.  
This discussion about the models is now completed by verifying the axioms.
\subsubsection{Axioms checking}
The classical axioms are implied since the model is a power set.
Other axioms are derived as follows.
\begin{property}[axioms verification]\label{property:model:2}\rien
\begin{description}
\item[$Partition$:]
$\modelalt{\bigvee_{p\in \mathbf{P}}p}=\modelalt{\top}$ and $\modelalt{p\wedge q}=\modelalt{\bot}=\emptyset\,,$ for $p,q\in\mathbf{P}$ such that $p\ne q$\,,
\item[$N$:]
$\modelalt{[J]\top}=\modelalt{\top}\,,$
\item[$K$:]
$\modelalt{[J](\neg X\vee Y)}\cap\modelalt{[J]X}\subseteq\modelalt{[J]Y}\,,$
for any $X,Y\in\mathcal{L}$\,,
\item[$Inc$:]
$\modelalt{[J](X)}\subseteq\modelalt{[J\cup K]X}\,,$
for any $X,Y\in\mathcal{L}$\,,
\item[$Ind$:]$\modelalt{[J\cup K]\setvee \YModel}\subset  \bigcup_{\mbox{\scriptsize$\begin{array}{@{}l@{}}F,G\subseteq\mathbf{P}\\F\cap G\subseteq \YModel\end{array}$}}\left(\modelalt{[J] \setvee F}\cap \modelalt{[K]\setvee G}\right)\;.$
\item[$T$:] $\modelalt{[J](X)}\subseteq\modelalt{X}$\,, when the model follows definition~(\ref{model:def:min:elt:withT}).
\end{description} 
\end{property}
Proofs indications are given in appendix~\ref{appendix:model:proof:1}.
\subsection{From logical combination to belief combination}
The interpretation of belief functions and belief combinations by means of modal propositions needs the concept of probability over logical propositions.   
\begin{definition}[probability on logical propositions]\label{definition:21}
Let $\equivclass{\mathcal{F}}\subset\equivclass{\mathcal{L}}$ be a finite set of propositions of $\equivclass{\mathcal{L}}$\,, closed under the classical operators (\emph{i.e.} $\rightarrow$ and $\bot$ and derived operators).
A probability distribution over $\equivclass{\mathcal{F}}$ is a mapping \mbox{$P:\equivclass{\mathcal{F}}\rightarrow \Rset$} such that:
$$P\ge0\;,\ \ P(\top)=1
\mbox{ \ and \ }
P(\equivclass{X})+P(\equivclass{Y})=P(\equivclass{X}\wedge \equivclass{Y})+P(\equivclass{X}\vee \equivclass{Y})\;.$$
\end{definition}
Now, it is assumed $s$ sensors characterized by the subsets $J_{1:s}\subseteq I$\,, such that $J_i\cap J_j=\emptyset$ for any $i\ne j$\,.
The logical framework $\equivclass{\mathcal{F}}$ is then constructed as a classical closure of the sets of simple \mbox{$[J_i]$-modalities}, \emph{i.e.}:
$$
\left\{[J_i]\equivclass{X}\:\big/\;i=1:s, \equivclass{X}\in\equivclass{\mathcal{L}_C}\right\}\subseteq\equivclass{\mathcal{F}}\;,.
$$
The set $\equivclass{\mathcal{F}}$ is a finite set.
It happens from previous properties that the propositions $\bigwedge_{i=1:s}(J_i)\equivclass{Y_i}$, for $\equivclass{Y_{1:s}}\in\equivclass{\mathcal{L}_C}$, constitute a partition generating $\equivclass{\mathcal{F}}$ and that $\left[\bigcup_{i=1:s}J_i\right]\equivclass{X}\in \equivclass{\mathcal{F}}$\,.
It is then possible to define the notion of belief, based on a probability defined over the framework $\equivclass{\mathcal{F}}$\,.
\begin{definition}[belief as probabilistic modality]\label{definition:22}
It is defined
$\bel_i(\equivclass{X})\stackrel{\Delta}{=}P([J_i]\equivclass{X})-P([J_i]\equivclass{\bot})$
and: $$\bel_1\oplusEMR\cdots\oplusEMR\bel_s(\equivclass{X})=P\left(\left[\bigcup_{i=1:s}J_i\right]\equivclass{X}\right)-P\left(\left[\bigcup_{i=1:s}J_i\right]\equivclass{\bot}\right)\;.$$
for $i=1:s$ and for any $\equivclass{X}\in\equivclass{\mathcal{L}_C}$\,.
\end{definition}
From this definition and the properties of $(J)$, especially property~(\ref{DSmTb2:Dmb:Sec:5:subsec:rule:eq:4:extended}), a definition of the bbas and of their combinations is recovered.
\begin{property}[logical definition of bba]\label{property:proba:1}
 $m_i(\equivclass{X})=P((J_i)\equivclass{X})$\,, and:
\begin{equation}
\label{DSmTb2:Dmb:Sec:5:subsec:logtobelfus:eq:3}
m_1\oplusEMR\cdots\oplusEMR m_s(\equivclass{X})=\sum_{\equivclass{Y_{1:s}}\in\equivclass{\mathcal{L}_C}:\bigwedge_{i=1:s}\equivclass{Y_i}= \equivclass{X}} P\left(\bigwedge_{i=1:s}(J_i)\equivclass{Y_i}\right)\;.
\end{equation}
\end{property}
In order to reduce~$P\left(\bigwedge_{i=1:s}(J_i)\equivclass{Y_i}\right)$\,, the independence of the \emph{disjoint} sources $J_{1:s}$ is instrumental in a probabilistic paradigm.
The idea is to compute $P\left(\bigwedge_{i=1:s}(J_i)\equivclass{Y_i}\right)$ by maximizing the entropy of $P$ over the logical framework $\equivclass{\mathcal{F}}$:
\begin{equation}
\label{DSmTb2:Dmb:Sec:5:subsec:logtobelfus:eq:4}
\begin{array}{@{}l@{}}\displaystyle
P\in\arg\max_{Q}\sum_{\equivclass{Y_{1:s}}\in\equivclass{\mathcal{L}_C}}-Q\left(\bigwedge_{i=1:s}(J_i)\equivclass{Y_i}\right) \ln Q\left(\bigwedge_{i=1:s}(J_i)\equivclass{Y_i}\right)
\;,
\vspace{7pt}\\\displaystyle
\mbox{u.c.}\ \ 
Q\left((J_i)\equivclass{Y_i}\right)=m_i((J_i)\equivclass{Y_i})\mbox{ for any }i=1:s\mbox{ and }\equivclass{Y_{1:s}}\in\equivclass{\mathcal{L}_C}\;.
\end{array}
\end{equation}
Combining~(\ref{DSmTb2:Dmb:Sec:5:subsec:logtobelfus:eq:3}) and~(\ref{DSmTb2:Dmb:Sec:5:subsec:logtobelfus:eq:4}), it becomes possible to derive \mbox{$m_1\oplusEMR\cdots\oplusEMR m_s$} from $m_{1:s}$\,.
Two cases arise, which are consequences of the model constructed in previous section:
\begin{itemize}
\item Axiom $T$ is removed.
Then, it is possible to have $\bigwedge_{i=1:s}(J_i)\equivclass{Y_i}\ne\bot$\, even if $\bigwedge_{i=1:s}Y_i=\bot$\,, as seen in the model construction.
As a consequence, the optimization implies the conjunctive rule. We are in the hypothesis of an \emph{open world.}
\item Axiom $T$ is used.
As seen in the model construction, this axiom reduces in the constraint~(\ref{def:model:modal:versionof:T}) for the simple modal propositions.
As a consequence, it is necessary to consider the constraint $\bigwedge_{i=1:s}(J_i)\equivclass{Y_i}=\bot$\,, when $\bigwedge_{i=1:s}\equivclass{Y_i}=\bot$\,.
This results in the following implied constraint in the optimization~(\ref{DSmTb2:Dmb:Sec:5:subsec:logtobelfus:eq:4}):
$$
\bigwedge_{i=1:s}\equivclass{Y_i}=\bot
\mbox{ implies }
Q\left(\bigwedge_{i=1:s}(J_i)\equivclass{Y_i}\right)=0\;.
$$
The rule EMR is thus derived, from a logical interpretation.
Moreover, we are able to distinguish between the open and closed world hypotheses.
From a modal point of view, the closed world (\emph{resp.} open world) hypothesis means that $[J]\bot\rightarrow \bot$ is necessary (\emph{resp.} is false in general).
From this modal interpretation, it appears, that the experts from an \emph{open world} may imply contradictions from factual information (typically because his knowledge of the real world is not exhaustive). 
\end{itemize}
\section{Conclusion}
\label{DSmTb2:Dmb:Sec:6}
A new combination rule have been defined for combining belief functions.
This rule is derived from a logical derivation, based on axioms of a multimodal logic, and from a principle of sources independence, based on the maximization of the system entropy.
This contribution implies a versatile framework for defining the combination rules and their implied intuition.
Indeed, it is possible to derive many other logics for defining the notion of belief and their combination at a logical level.
\\[5pt]
In this contribution especially, we proposed a logic with strong but desirable properties, and we derived some conlusions about the feasibility of the related combinations.
These properties are typically the conservation of belief during the combination, and the compliance with a probabilistic ground truth.
It appeared that such a rule is theoretically constructible.
But however, it appeared that as soon as the principle of independence is used, it introduces information which imply a bias which is conflicting in general with the ground truth.
This information should be used at the final stage of the fusion, which is a strong constraint on the fusion process. 
We then questioned about the need to release the principle of belief conservation (as do most rules with conflict redistribution) or to abandon the compliance with a ground truth reference.
We also suggested that these issues may be solved by using conditional belief functions more thoroughly  instead of absolute belief functions.
%
%
%

\small
\appendix

\section{Proofs of section~\ref{DSmTb2:Dmb:Sec:4}}
\label{appendix:proof:1}
For the sake of simplicity, somes proofs are done for $s=1$ or $s=2$.
But these results are easily generalized to $s\ge 1$ sources.
%
\paragraph{Belief sharpening -- property \ref{property:1} --}$\lhd$ is an order relation.
\begin{proof}\rien\\
\emph{Proof of transitivity.}
Let $ r_{13}$ be defined according to~(\ref{dmb:EMR:mathbbT:eq:1}).
\\
Then, it is clear that $ r_{13}\ge0$.
\\
It also comes:
$$\begin{array}{@{}l@{}}\displaystyle
\sum_{X_1\in G^\Theta} r_{13}(X_1,X_3)
=
\sum_{X_2\in G^\Theta}\frac{ r_{23}(X_2,X_3)}{m_2(X_2)}\sum_{X_1\in G^\Theta} r_{12}(X_1,X_2)
=
\vspace{5pt}\\\displaystyle
\rien\qquad\qquad
\sum_{X_2\in G^\Theta}\frac{ r_{23}(X_2,X_3)}{m_2(X_2)}m_2(X_2)
=\sum_{X_2\in G^\Theta} r_{23}(X_2,X_3)=m_3(X_3)\;,
\end{array}$$
and
$$\begin{array}{@{}l@{}}\displaystyle
\sum_{X_3\in G^\Theta} r_{13}(X_1,X_3)
=
\sum_{X_2\in G^\Theta}\frac{ r_{12}(X_1,X_2)}{m_2(X_2)}\sum_{X_3\in G^\Theta} r_{23}(X_2,X_3)
=
\vspace{5pt}\\\displaystyle
\rien\qquad\qquad
\sum_{X_2\in G^\Theta}\frac{ r_{12}(X_1,X_2)}{m_2(X_2)}m_2(X_2)
=\sum_{X_2\in G^\Theta} r_{12}(X_1,X_2)=m_1(X_1)\;.
\end{array}$$
Now, since $ r_{12}(X_1,X_2)=0$ for any $X_1,X_2\in G^\Theta$ such that $X_2\not\subseteq X_1$, and $ r_{23}(X_2,X_3)=0$ for any $X_2,X_3\in G^\Theta$ such that $X_3\not\subseteq X_2$, it comes that $ r_{12}(X_1,X_2) r_{23}(X_2,X_3)=0$ for any $X_1,X_2,X_3\in G^\Theta$ such that $X_3\not\subseteq X_1$\,.
As a consequence:
$$
 r_{13}(X_1,X_3)=0\mbox{ \ for any \ }X_1,X_3\in G^\Theta\mbox{ such that } X_3\not\subseteq X_1\;,
$$
which also results in:
$$
\sum_{X_1\in G^\Theta} r_{13}(X_1,X_3)=\sum_{X_1:X_3\subseteq X_1} r_{13}(X_1,X_3)
\;,
$$
and
$$
\sum_{X_3\in G^\Theta} r_{13}(X_1,X_3)=\sum_{X_3:X_3\subseteq X_1} r_{13}(X_1,X_3)
\;.
$$
Then, $ r_{13}$ is indeed an sharpening from $\bel_{1}$ to $\bel_{3}$\,.
\rien\\[5pt]
\emph{Proof of reflexivity.}
\\
The reflexivity itself is obvious, since $ r$, defined by $ r(X,X)=m(X)$ and $ r(X,Y)=0$ for any $X,Y\in G^\Theta$ such that $X\ne Y$, will check all requirements for a sharpening from $\bel$ to itself.
\\
Now, assume another sharpening $ r'$ from $\bel$ to itself.
Then, there are propositions $X\in G^\Theta$ such that $ r'(X,X)<m(X)$\,.
Let $X_M\in G^\Theta$ be a maximal proposition, in regards to the order $\subseteq$, such that $ r'(X_M,X_M)<m(X_M)$\,.
It comes from this definition:
$$
 r'(X_M,X_M)<m(X_M)
\mbox{ \ and \ }
 r'(X,X_M)=0\;,
$$
for any $X\in G^\Theta\setminus\{X_M\}$ such that $X_M\subset X\,.$
As a consequence:
$$
\sum_{X\in G^\Theta}  r'(X,X_M)<m(X_M)\;,
$$
which is a contradiction with the hypothesis.
\rien\\[5pt]
\emph{Proof of antisymmetry.}
\\
It comes from the transitivity that:
$$
 r_{11}(X_1,Y_1)=\sum_{X_2\in G^\Theta}\frac{ r_{12}(X_1,X_2) r_{21}(X_2,Y_1)}{m_2(X_2)}
$$
is a sharpening from $\bel_1$ to $\bel_1$\,.
From the unicity of this sharpening, it comes:
\begin{equation}
\label{lemma:1:eq:1}
\sum_{X_2\in G^\Theta}\frac{ r_{12}(X_1,X_2) r_{21}(X_2,X_1)}{m_2(X_2)}= m_1(X_1)\;,
\end{equation}
and
\begin{equation}
\label{lemma:1:eq:2}
\sum_{X_2\in G^\Theta}\frac{ r_{12}(X_1,X_2) r_{21}(X_2,Y_1)}{m_2(X_2)}=0\;,
\end{equation}
for any $X_1\in G^\Theta$ and $Y_1\in G^\Theta\setminus\{X_1\}$\,.
Similarly, it is obtained:
\begin{equation}
\label{lemma:1:eq:3}
\sum_{X_1\in G^\Theta}\frac{ r_{21}(X_2,X_1) r_{12}(X_1,X_2)}{m_1(X_1)}= m_2(X_2)\;,
\end{equation}
for any $X_2\in G^\Theta$\,.
\\[3pt]
Since $ r_{12}(X_1,X_2)=0$ or $ r_{21}(X_2,X_1)=0$ when $X_1\ne X_2$, it comes from equations (\ref{lemma:1:eq:1}) and (\ref{lemma:1:eq:3}):
$$
\frac{ r_{12}(X_1,X_1) r_{21}(X_1,X_1)}{m_2(X_1)}=m_1(X_1)\mbox{ \ for any }X_1\in G^\Theta\;,
$$
and
$$
\frac{ r_{21}(X_2,X_2) r_{12}(X_2,X_2)}{m_1(X_2)}=m_2(X_2)\mbox{ \ for any }X_2\in G^\Theta\;.
$$
As a corollary, and since $ r_{ij}\le m_i$ and  $ r_{ij}\le m_j$ , it comes:
$$
m_2(X)\ne0 \iff m_1(X)\ne0\iff  r_{12}(X,X)\ne0 \iff  r_{21}(X,X)\ne0
\;.
$$
Since $ r_{12}\ge0$ and $ r_{21}\ge0$\,, it comes from equation~(\ref{lemma:1:eq:2}):
$$
 r_{12}(X_1,X_2)=0\mbox{ \ or \ } r_{21}(X_2,Y_1)=0\;,
$$
for any $X_1, X_2\in G^\Theta$ and $Y_1\in G^\Theta\setminus\{X_1\}$\,.
By applying this result to $X_2=Y_1$ and $X_2\ne X_1$, it is derived $ r_{12}(X_1,X_2)=0$ when $m_2(X_2)\ne0$\,.
Since anyway $0\le r_{12}(X_1,X_2)\le m_2(X_2)$, it follows the result:
$$
 r_{12}(X_1,X_2)=0\mbox{ for any }X_1,X_2\in G^\Theta\mbox{ such that }X_1\ne X_2\;.
$$
Similarly, it may be proved:
$$
 r_{21}(X_2,X_1)=0\mbox{ for any }X_1,X_2\in G^\Theta\mbox{ such that }X_1\ne X_2\;.
$$
As a consequence:
$$\begin{array}{@{}l@{}}\displaystyle
 r_{12}(X,X)=\sum_{X_1\in G\Theta} r_{12}(X_1,X)=m_2(X)=\sum_{X_1\in G\Theta} r_{21}(X,X_1)
\vspace{5pt}\\\displaystyle
\rien\qquad\qquad= r_{21}(X,X)=\sum_{X_2\in G\Theta} r_{21}(X_2,X)=m_1(X)\;.
\end{array}$$
\end{proof}
\paragraph{Belief sharpening -- property \ref{property:2} --}
$\bel_1\lhd\bel_2$ implies $\bel_1\le\bel_2$\,.
\begin{proof}Let $ r_{12}$ be a sharpening from $\bel_1$ to $\bel_2$\,.
Since:
$$
\bigl\{(Y_1,Y_2)
\;\big/\;Y_2\subseteq Y_1\subseteq X\bigr\}
\subseteq
\bigl\{(Y_1,Y_2)
\;\big/\;Y_2\subseteq Y_1\mbox{ and } Y_2\subseteq X\bigr\}\;,
$$
it comes:
$$\mbox{\small{\rien\hspace{-30pt}$\displaystyle
\bel_1(X)=\sum_{Y_1:Y_1\subseteq X}\ \sum_{Y_2:Y_2\subseteq Y_1} r_{12}(Y_1,Y_2)
\le
\sum_{Y_2:Y_2\subseteq X}\ \sum_{Y_1:Y_2\subseteq Y_1} r_{12}(Y_1,Y_2)=\bel_2(X)\;.
$}}$$
\end{proof}
%
%
\label{DSmTb2:Dmb:Sec:4:subsec:prop:appendix}
%
%
\paragraph{Neutral element -- property~\ref{property:5} --}
$m\oplusEMR\nu$ exists, and $m\oplusEMR\nu=m$\,.
\begin{proof}
Since $\sum_Y f_o(Y,Z)=\nu(Z)$ and $f_o\ge0$\,, it is deduced $f_o(Y,Z)=0$ unless $Z=\everTrue$.
Now, since $\sum_Z f_o(Y,Z)=m(Y)$, it is concluded:
$$
f_o(Y,\everTrue)=m(Y)
\quad\mbox{and}\quad
f_o(Y,Z)=0\mbox{ for }Z\ne\everTrue\;.
$$
Since this function obviously satisfies the constraints, the existence of $m\oplusEMR\nu$ is implied.
Then the result $m\oplusEMR\nu=m$ is immediate.
\end{proof}
%
\paragraph{Conservation of belief -- property~\ref{property:6} --}
$\bel_1\;\lhd\;\bel_1\oplusEMR\bel_2$\,.
\begin{proof}By definition, $m_1\oplusEMR m_2(X)= \sum_{V,W:V\cap W= X}f_o(V,W)$\,, where $f_o$ is an optimal solution to~(\ref{dmb:dsmtb2:ch2:rule:eq:5}).
\\
As a consequence, $m_1\oplusEMR m_2(X)= \sum_{V} r(V,X)\,,$ where:
$$
 r(V,X)=\sum_{W: V\cap W=X}f_o(V,W)\;.
$$ 
Now $ r$ verifies $ r(V,X)=0$ for $X\not\subset V$ and, since $\sum_Wf_o(V,W)=m_1(V)$, it comes:
$$
\sum_X r(V,X)=\sum_{X,W: V\cap W=X}f_o(V,W)=\sum_{W}f_o(V,W)=m_1(V)\;.
$$
\end{proof}
Corollaries are obvious.
%
%
\paragraph{Probabilistic bound -- property~\ref{property:10} --}
\begin{proof}
The existence of $\bel_1\oplusEMR\cdots\oplusEMR\bel_s\oplusEMR P$ is a consequence of the combination of dominated belief function, since, indeed, $P\lhd P$\,.
\\[5pt]
From the conservation of belief, it happens:
$$
P\lhd \bel_1\oplusEMR\cdots\oplusEMR\bel_s\oplusEMR P\;.
$$
Since $P$ is a probability measure, it is minimal and the property is implied.
\end{proof}
%
\paragraph{Idempotence -- property~\ref{property:13} --}
$m\emph{ is probabilistic}\Rightarrow m\oplusEMR \cdots \oplusEMR m=m\,.$
\begin{proof}Since $m$ is probabilistic, there are $X_{1:n}\in G^\Theta\setminus\{\everFalse\}$ such that:
$$
\sum_{j=1}^n m(X_j)=1\mbox{ \ and \ }
\left\{\begin{array}{l@{}}
i\ne j\Rightarrow X_i\cap X_j=\everFalse\;,
\vspace{4pt}\\
\bigcup_{j=1}^n X_j=\everTrue\;.
\end{array}\right.$$
As a consequence, the only function $f$, which fulfills the constraint of~(\ref{dmb:dsmtb2:ch2:rule:eq:5}), is defined by:
$$
f(X,\cdots,X)=m(X)\mbox{ \ for any }X\in G^\Theta\;.
$$
Of this, the property is an immediate consequence.
\end{proof}
\section{Proofs of section~\ref{logpart:sect:3} and~\ref{logpart:sect:2}}
\label{appendix:proof:2}
%
%
\label{appendix:modal:proof:1}
\paragraph{Bla exclusivity -- property~\ref{property:25} --}
\begin{proof}
It is noticed that $E\cap F\subsetneq E$ or $E\cap F\subsetneq F$ when $E\neq F$\,.
Then, the proof is achieved by deriving:
$$
(J)\setvee E\wedge(J)\setvee{F}\equiv[J]\setvee{E}\wedge[J]\setvee{F}\wedge\left(
\bigwedge_{G\subsetneq E}\neg[J]\setvee G\right)\wedge\left(
\bigwedge_{H\subsetneq F}\neg[J]\setvee H\right)\;,
$$
and combining it with $[J]\setvee{E}\wedge[J]\setvee{F}\equiv[J](\setvee{E}\wedge\setvee{F}) \equiv [J]\setvee{(E\cap F)}$\,.
\end{proof}
\paragraph{Bla exhaustivity -- property~\ref{property:26} --}
\begin{proof}
The proof is recursive.
\\
It is first noticed that $[J]\setvee\emptyset=(J)\setvee\emptyset$\,.
\\[5pt]
Now, let $E\subseteq \mathbf{P}$ and assume $\bigvee_{G\subseteq F}
(J)\setvee G=[J]\setvee F$ for any $F\subsetneq E$\,.
\\
Then:
$$
\bigvee_{F\subseteq E}(J)\setvee F\equiv
(J)\setvee E \vee\left(\bigvee_{F\subsetneq E}\;\;\bigvee_{G\subseteq F}(J)\setvee G\right)\equiv
(J)\setvee E \vee\left(\bigvee_{F\subsetneq E}[J]\setvee F\right)\,.
$$
It follows $\bigvee_{F\subseteq E}(J)\setvee F\equiv[J]\setvee E\vee\left(\bigvee_{F\subsetneq E}[J]\setvee F\right)\,,$ from the difinition of $(J)\setvee E$.
\\
Since the modality is non-decreasing, it is deduced $\bigvee_{F\subseteq E}(J)\setvee F\equiv[J]\setvee E$\,.
\end{proof}
%

%
%
\paragraph{Computing the combination -- property~\ref{property:30} and lemma~\ref{property:32} --}
\begin{proof}[Proof of lemma]
From~(\ref{F2K7:EMR:small:prop:4}), (\ref{F2K7:EMR:small:prop:1}) and axiom $Ind$, applied on $J\cup K$, it comes:
$$\bigvee_{F\cap G\subseteq E}([J]\setvee F\wedge[K]\setvee G)\equiv[J\cup K]\setvee E\;.$$
Then, by $[J]\setvee F\equiv\bigvee_{F_2\subseteq F}(J)\setvee F_2$
and $[K]\setvee G\equiv\bigvee_{G_2\subseteq G}(K)\setvee G_2$\,, it comes:
$$\rien\hspace{-30pt}
[J\cup K]E\equiv\hspace{-5pt}\bigvee_{F\cap G\subseteq E}\;\;
\bigvee_{F_2\subseteq F}\;\;\bigvee_{G_2\subseteq G} \left((J)\setvee F_2\wedge(K)\setvee G_2\right)
\equiv\hspace{-5pt}\bigvee_{F_2\cap G_2\subseteq E} \hspace{-5pt}\left((J)\setvee F_2\wedge(K)\setvee G_2\right)\;.
$$
\end{proof}
\begin{proof}[Main proof]
From the definition of the bla and lemma, it is deduced:
$$
(J\cup K)E\equiv\bigvee_{F\cap G\subseteq E}\left((J)\setvee F\wedge(K)\setvee G\right)
\wedge\neg
\bigvee_{F\cap G\subsetneq E}\left((J)\setvee F\wedge(K)\setvee G\right)
\;.
$$
The propositions $(J)\setvee F\wedge(K)\setvee G$ constituting a partition, it comes:
$$
(J\cup K)E\equiv \bigvee_{F\cap G= E}\left((J)\setvee F\wedge(K)\setvee G\right)
\;.
$$
\end{proof}
%
%
\label{appendix:model:proof:1}

\paragraph{Axioms verification -- property~\ref{property:model:2} --}
\begin{proof}[Proof sketch]
\emph{Partition}, $K$, $N$ and $Inc$ are almost immediate from the definition.
Axiom $K$ is derived from the definition of $\modelalt{[J]\cdots}$ and from the property:
$$
\modelof{\bigcap_{j\in J}\YModel_j}\subseteq(\modelalt{\top}\setminus\modelalt{X})\cup\modelalt{Y}
\mbox{ and }\modelof{\bigcap_{j\in J}\YModel_j}\subseteq\modelalt{X}
\mbox{ \ imply \ }
\modelof{\bigcap_{j\in J}\YModel_j}\subseteq\modelalt{X}\cap\modelalt{Y}\subseteq\modelalt{Y}
\;.
$$
Axiom $Ind$ follows from the following derivation:
$$\begin{array}{@{}l@{}}\displaystyle
\modelalt{[J\cup K]\setvee E}=
\bigcup_{\xModel_0\in\mathbf{P}}
\bigcup_{\YModel_{I\setminus (J\cup K)}}
\bigcup_{\YModel_{J\cup K}:\modelof{\bigcap_{j\in J\cup K}\YModel_j}\subseteq\modelof{E}}
\phi(\xModel_0,\YModel_{1:n})
\vspace{4pt}\\\displaystyle
\rien\qquad\qquad=
\bigcup_{\xModel_0\in\mathbf{P}}
\bigcup_{\YModel_{I\setminus (J\cup K)}}
\bigcup_{F\cap G\subseteq E}
\bigcup_{\YModel_{J\cup K}:
\mbox{\scriptsize$\left\{\begin{array}{@{}l@{}}
\modelof{\bigcap_{j\in J}\YModel_j}\subseteq\modelof{F}
\\
\modelof{\bigcap_{j\in K}\YModel_j}\subseteq\modelof{G}
\end{array}\right.$}
}
\phi(\xModel_0,\YModel_{1:n})
\;.\end{array}
$$
At last, axiom $T$ is derived from the following derivation based on hypothesis~(\ref{model:def:min:elt:withT}):
$$
\begin{array}{@{}l@{}}\displaystyle
\modelalt{[J]\setvee E}=
\bigcup_{\xModel_0\in\mathbf{P}}
\bigcup_{\YModel_{I\setminus J}}
\bigcup_{\YModel_{J}:\modelof{\bigcap_{j\in J}\YModel_j}\subseteq\modelof{E}}
\phi(\xModel_0,\YModel_{1:n})
=
\bigcup_{\xModel_0\in E}
\bigcup_{\YModel_{I\setminus J}}
\bigcup_{\YModel_{J}:\modelof{\bigcap_{j\in J}\YModel_j}\subseteq\modelof{E}}
\phi(\xModel_0,\YModel_{1:n})
\vspace{4pt}\\\displaystyle
\rien\qquad\qquad\subseteq
\bigcup_{\xModel_0\in E}
\bigcup_{\YModel_{I}}
\phi(\xModel_0,\YModel_{1:n})
=
\modelalt{\setvee E}\;,
\end{array}$$ 
and from the property $\modelalt{[J]X}=\modelalt{[J]\setvee\pi(X)}$\,, where $\pi(X)=\bigvee_{\YModel\subseteq\mathbf{P}:\modelof{\YModel}\subseteq\modelalt{X}}E$\,.
\end{proof}

\begin{thebibliography}{99}
%
\bibitem{blackburn}
Blackburn P., De Rijke M., Venema Y.,
\emph{Modal Logic (Cambridge Tracts in Theoretical Computer Science)},
Cambridge University Press, 2002.
%
\bibitem{Dempster1967}
Dempster A.P., \emph{Upper and Lower probabilities induced by a multivalued
mapping}, Annals of Mathematical Statistics, vol. 83, pp. 325--339, 1967.
%
\bibitem{dempster2}
Dempster A.~P., \emph{A generalization of Bayesian inference}, J. Roy. Statist. Soc., Vol. B, No. 30, pp. 205--247, 1968.
%
%
\bibitem{Shafer1976}
Shafer G., \emph{A mathematical theory of evidence}, Princeton
University Press, 1976.
%
%
\bibitem{Denoeux2006}
Denoeux T., \emph{The cautious rule of combination for belief functions
and some extensions}, International Conference on Information Fusion,
Florence, Italy, 10--13 July 2006.
%
\bibitem{Dubois86} 
D.~Dubois and H.~Prade, ``On the unicity of Dempster rule of combination,'' \emph{International Journal of Intelligent Systems}, vol.~1, No.~2, pp. 133-142, 1986.
%
\bibitem{Dubois88} 
D.~Dubois and H.~Prade, ``Representation and Combination of uncertainty with belief functions and possibility measures,'' \emph{Computational Intelligence}, vol.~4, pp. 244-264, 1988.
%
\bibitem{Florea2006}
Florea M.C., Dezert J., Valin P., Smarandache F. and Jousselme A.L.,
\emph{Adaptative combination rule and proportional conflict redistribution rule
for information fusion}, COGnitive systems with Interactive Sensors,
Paris, France, March 2006
%
\bibitem{Lefevre2002}
Lefevre E., Colot O., Vannoorenberghe P., \emph{Belief functions combination
and conflict management}, Information Fusion Journal, Elsevier Publisher,
Vol. 3, No. 2, pp. 149--162, 2002.
%
\bibitem{martin2007}
Martin A. and Osswald C., \emph{Toward a combination rule to deal with partial
conflict and specificity in belief functions theory}, International Conference
on Information Fusion, Qu\'ebec, Canada, 9--12 July 2007.
%
\bibitem{Martin2006}
Martin A. and Osswald C., \emph{A new generalization of the proportional
conflict redistribution rule stable in terms of decision}, Applications and
Advances of DSmT for Information Fusion, Book 2, American Research
Press Rehoboth, F. Smarandache and J. Dezert, pp. 69--88 2006.
%
\bibitem{martin2}
Osswald C., Martin A.,
\emph{Understanding the large family of Dempster-Shafer theory's fusion operators - a decision-based measure},
9th international conference on information fusion, Florence, 2006.

\bibitem{Smarandache2005}
Smarandache F. and Dezert J., \emph{Information Fusion Based on New
Proportional Conflict Redistribution Rules}, International Conference on
Information Fusion, Philadelphia, USA, 25--29 July 2005.
%
\bibitem{Smarandache2006}
Smarandache F., Dezert J., \emph{Proportional Conflict Redistribution Rules for
Information Fusion},  Applications and Advances of DSmT for Information Fusion, Book 2, American Research Press Rehoboth, F. Smarandache and J. Dezert, pp. 3--68, 2006.
%
\bibitem{smets}
Smets Ph., \emph{The combination of evidences in the transferable belief model}, IEEE Transactions on Pattern Analysis and Machine Intelligence, Vol. 12, no. 5, pp. 447--458, 1990.
%
\bibitem{Smets07}
Ph.~Smets, ``Analyzing the combination of conflicting belief functions,'' \emph{Information Fusion}, vol.~8, no.~4, pp. 387-412, 2007.
%
\bibitem{Yager1987}
Yager R.R., \emph{On the Dempster-Shafer Framework and New Combination
Rules}, Informations Sciences, vol. 41, pp. 93--137, 1987.
%
\bibitem{DSmTb1:dmb}
Dambreville F.,
\emph{Probabilistic logics related to DSmT}
in \emph{Advances and Applications of DSmT for Information Fusion}, F. Smarandache and J. Dezert editors, American Research Press, Rohoboth, 2004.
%
\bibitem{DSmTb2:dmb}
Dambreville F.,
\emph{Conflict Free Rule for Combining Evidences}
in \emph{Advances and Applications of DSmT for Information Fusion, Vol. 2}, F. Smarandache and J. Dezert editors, American Research Press, Rohoboth, 2006.
%
\bibitem{F2K7:dmb:emr}
Dambreville F., \emph{Combining Evidences by means of the Entropy Maximization Principle},  International Conference on
Information Fusion, Quebec, Canada, july 2007.
%
\bibitem{DSmTBook1}
F. Smarandache and J. Dezert editors, \emph{Advances and Applications of DSmT for Information Fusion}, American Research Press, Rohoboth, 2004.
%
\bibitem{DSmTBook3_chap6}
Fr\'ed\'eric Dambreville, \emph{Chap. 6: Definition of evidence fusion rules based on referee functions}, in Smarandache F. \& Dezert J., Editors, \emph{Applications and Advances on DSmT for Information Fusion (Collected Works)}, Vol. 3, American Research Press, 2009.
%
%
%
%
\bibitem{besnard1996}
Besnard P., Jaouen P., Perin J. Ph., \emph{Extending the transferable belief model for inconsistency handling}, Information Processing and Management of Uncertainty, 1996.
%
%
\bibitem{DSmTBook2}
Smarandache F., Dezert J. (Editors), \emph{Applications and Advances on DSmT for Information Fusion (Collected Works)}, Vol. 2, American Research Press, June 2006.

%
\bibitem{dezert}
Dezert J., \emph{Foundations for a new theory of plausible and paradoxical reasoning}, Information \& Security, An international Journal, edited by Prof. Tzv. Semerdjiev, CLPP, Bulg. Acad. of Sciences, Vol. 9, 2002.

\bibitem{ruspini1987}
Ruspini E., \emph{Epistemic logics, probability, and the calculus of evidence}, In Proceedings of the 10th International Joint Conference on Artificial Intelligence, 1987.
%


 \end{thebibliography}
\end{document}